\documentclass{article}

\usepackage[utf8]{inputenc}
\usepackage[dvipsnames]{xcolor}
\usepackage{mathtools}
\usepackage{url}
\usepackage{graphicx}
\usepackage{latexsym}
\usepackage{xspace}
\usepackage{amsmath}
\usepackage{amssymb}
\usepackage{amsthm}
\usepackage{mathrsfs}

\usepackage{kr}
\usepackage{times}
\usepackage{soul}
\usepackage[hidelinks]{hyperref}
\usepackage[small]{caption}
\usepackage{booktabs}

\usepackage[ruled,vlined,linesnumbered]{algorithm2e}
\usepackage{tikz}
\usepackage{soul}
\usetikzlibrary{graphs,quotes,bending,arrows.meta,positioning} 

\newtheorem{example}{Example}
\newtheorem{theorem}{Theorem}
\newtheorem{proposition}{Proposition}
\newtheorem{definition}{Definition}
\newtheorem{lemma}{Lemma}

\newtheorem{corollary}{Corollary}

\definecolor{darkgray}{rgb}{0.2, 0.2, 0.2}

\author{%
Michel Leclère$^1$\and
Marie-Laure Mugnier$^1$\and
Guillaume Pérution-Kihli$^{1}$\\
\affiliations
$^1$LIRMM, Inria, University of Montpellier, CNRS, France 
\emails
\{leclere,mugnier\}@lirmm.fr,
guillaume.perution-kihli@inria.fr
}

\newcommand{\mlmm}[1]{{\color{black}{#1}}} 

\newcommand{\mic}[1]{{\color{black}{#1}}}

\definecolor{cssgreen}{rgb}{0.0, 0.5, 0.0} 
\definecolor{forestgreen}{rgb}{0.0, 0.54, 0.26} 

  \newcommand{\fb}{F} 
  \newcommand{\fr}[1]{\mathtt{fr}(#1)} 
  \newcommand{\exist}[1]{\mathtt{exist}(#1)}
  \newcommand{\head}[1]{\mathtt{head}(#1)} 
  \newcommand{\headi}[2]{\mathtt{head}_{#1}(#2)} 
  \newcommand{\body}[1]{\mathtt{body}(#1)} 
  \newcommand{\vars}[1]{\mathtt{vars}(#1)} 
  \newcommand{\terms}[1]{\mathtt{terms}(#1)} 
  \newcommand{\consts}[1]{\mathtt{consts}(#1)} 
  \newcommand{\chase}[2]{\mathtt{chase}(#1,#2)} 
  
  \newcommand{\generate}[1]{\mathtt{generate}{#1}} 
  \newcommand{\cover}[1]{\mathtt{cover}{#1}} 
  \newcommand{\removeMoreSpecific}[1]{\mathtt{removeMoreSpecific}{#1}}

\newcommand\m{\mathcal{M}}

\newcommand{\vect}[1]{\mathbf{#1}}

\newcommand{\qUCQ}{\mathcal Q}
\newcommand{\qCQ}{Q}

\newcommand{\ruleset}{\mathcal{R}}
\newcommand{\homo}{h}

\title{Query Rewriting with Disjunctive Existential Rules and Mappings}

\begin{document}

\maketitle

\begin{abstract}
We consider the issue of answering unions of conjunctive queries (UCQs) with disjunctive existential rules and mappings.
While this issue has already been well studied from a chase perspective, query rewriting within UCQs has hardly been addressed yet. We first propose a sound and complete query rewriting operator, which has the advantage of establishing a tight relationship between a chase step and a rewriting step. The associated breadth-first query rewriting algorithm outputs a minimal UCQ-rewriting when one exists. 
\mlmm{Second, we show that for any ``truly disjunctive'' nonrecursive rule, there exists a conjunctive query that has no UCQ-rewriting.} 
It follows that the notion of finite unification sets (fus), which denotes sets of existential rules such that any UCQ admits a UCQ-rewriting, seems to have little relevance in this setting. Finally, turning our attention to mappings, we show that the problem of determining whether a UCQ admits a UCQ-rewriting through a disjunctive mapping is undecidable. We conclude with a number of open problems.

\medskip
This report contains the paper accepted at KR 2023 and an appendix with full proofs. 
\end{abstract}

\section{Introduction}

Existential rules  \cite{DBLP:conf/kr/CaliGK08,blms:09,Cali2009}, aka tuple generating dependencies \cite{BeeriVardi84},
 are an extension of datalog (i.e., first-order function-free Horn rules), which allows for existentially quantified variables in the rule heads, e.g., $\forall x(\text{human}(x) \rightarrow \exists y~\text{isParent}(y,x))$. 
They have become a popular language to model ontologies and do reasoning on data. 
Then, a key issue is \emph{ontology-mediated query answering}, which consists of computing the answers to a query on a knowledge base (KB), composed of a set of facts (or data) $F$ and an ontology $\mathcal O$. In this context, most works focus on the prominent class of (unions of) conjunctive queries ((U)CQs). 
There are two main dual techniques to compute the answers to a query $Q$: the \emph{chase}, which enriches the facts $F$ by performing a fixpoint computation with the ontology $\mathcal{O}$ until a canonical model of $F$ and $\mathcal{O}$ is obtained (then $Q$ is evaluated on this canonical model), and \emph{query rewriting}, where $Q$ is rewritten using $\mathcal{O}$ into a query $Q'$, such that for any set of facts $F$, the evaluation of $Q'$ on $F$ yields the answers to $Q$ on the KB. 
Query answering with general existential rules is undecidable, however a wide range of decidable subclasses have been defined,  
 based on syntactic restrictions that ensure the termination of chase-like or query rewriting techniques. 
Tuple generating dependencies (TGDs) are also the main formalism to represent \emph{schema mappings}, which are high-level specifications of the relationships between two database schemas \cite{DBLP:journals/tcs/FaginKMP05}. Schema mappings are at the core of many data interoperability tasks, such as data exchange, data integration or peer data management. 
More specifically, a mapping is a set of TGDs, with bodies and heads expressed on disjoint sets of predicates, namely $\mathcal S$ and $\mathcal T$, 
called the \emph{source} and the \emph{target} schemas.  
Given a database instance $I$ on $\mathcal S$ and a mapping $\mathcal M$, a query expressed on $\mathcal T$ is posed on the set of facts produced from $I$ by triggering  $\mathcal M$; again, query answering can be solved by \emph{chasing}  $I$ with $\mathcal M$ or \emph{rewriting} $Q$ with $\mathcal M$ into a query that is evaluated on $I$. 
Since mappings are inherently nonrecursive, both techniques always terminate. 
Finally, in the Ontology-Based Data Access \mlmm{(OBDA)} framework \cite{PoggiLCGLR08}, mappings specify relationships between a database schema and an ontology. Here, existential rules can be used as a uniform language to express both the ontology and the mapping \cite{DBLP:conf/kr/BuronMT21}.  
%

Existential rules generalize popular description logics (DLs) used to do reasoning on data, such as DL-Lite \cite{dl-lite}, $\mathcal{EL}$ \cite{DBLP:conf/ijcai/BaaderBL05,lutz:09}  and more expressive Horn-DLs \cite{KRH:HornSHIQ06}.   
However, they do not capture \emph{nondeterministic} features, as offered by some key DLs such as 
$\mathcal{ALC}$ \cite{DBLP:journals/ai/Schmidt-SchaussS91} or the Semantic Web ontology language OWL \cite{owl2-overview}. 

In this paper, we consider the extension of existential rules with \emph{disjunction}, 
%
e.g., $\forall x \forall y(\text{isGrandParent}(x,y) \rightarrow \exists z_1~(\text{isParent}(x,z_1) \land \text{isMother}(z_1,y)) \lor  \exists z_2~(\text{isParent}(x,z_2) \land \text{isFather}(z_2,y)))$.  From a KR perspective, the usefulness of such rules has long been acknowledged for ontology modeling, but also for expressing nondeterministic guessing in problem solving, see e.g., \cite{DBLP:journals/tods/EiterGM97}. 
From a database perspective, disjunction in schema mappings received considerable attention 
in the context of \emph{mapping management}, \mlmm{where mapping composition and inversion emerged as fundamental operators} 
\cite{DBLP:conf/vldb/BernsteinH07,DBLP:conf/pods/ArenasPRR10}. 
Indeed, disjunction is required to express several kinds 
\mlmm {of inverse mappings, like so-called quasi-inverses or maximum recovery mappings} \cite{DBLP:journals/tods/FaginKPT08,DBLP:conf/pods/ArenasPR08}. Beside the issue of constructing such mappings, the design of associated query answering techniques is highly relevant. For instance, in a peer data management system, a mapping $\mathcal M$ from peer $P_1$ to peer $P_2$ allows to rewrite a query on $P_2$ in terms of $P_1$, while an inverse of $\mathcal M$ allows to rewrite a query on $P_1$ in terms of $P_2$. As another example, consider a mapping $\m$ from schema $A$ to schema $B$, and assume that $A$ evolves into $A'$, which is expressed by a mapping $\m'$; the relation between $A'$ and $B$ can be obtained by inverting $\m'$ and composing it with $\m$; then, a query on $B$ can be translated into a query on $A'$ by rewriting it first with $\mathcal M$, then with the inverse of  $\m'$ 
\cite{DBLP:conf/dagstuhl/Perez13}. \mlmm{Such scenario is also relevant in OBDA, taking for $B$ an ontology instead of a schema.}

So far, reasoning with disjunctive existential rules has been mainly studied through the \emph{chase}. It was shown that decidable classes of (conjunctive) existential rules, based on the behavior of the chase, can be generalized to disjunctive rules in a quite natural way, whether in relation to acyclicity notions \cite{CARRAL17} or based on guardedness \cite{DBLP:journals/tplp/AlvianoFLM12,DBLP:conf/mfcs/GottlobMMP12,BOURHIS16}, although these generalizations come with a huge increase in the complexity of query answering.   

\mlmm{
In contrast, \emph{query rewriting} within UCQs has been barely addressed yet. 
A notable exception is the work in \cite{ALFONSO21}, which provides a rewriting technique based on first-order resolution (see Section \ref{sec-algorithm_ucq_rewriting}).
A large body of work has studied the rewritability of ontology-mediated queries, i.e., pairs of the form $(Q, \mathcal O)$ with $Q$ a (U)CQ and $\mathcal O$ an ontology, into query languages of various expressivity. However, for ontologies expressed in fragments of disjunctive existential rules, most studies target expressive rewriting languages, like disjunctive datalog \cite{DBLP:journals/tods/BienvenuCLW14,DBLP:conf/icdt/AhmetajOS18}. 
As far as we are aware, the only result directly relevant to our purpose comes from the fine-grained complexity study in \cite{DBLP:conf/kr/GerasimovaKKPZ20}, which provides syntactic rewritability conditions for ontology-mediated queries where the ontology is composed of a single specific disjunctive rule, called a covering axiom (see Section \ref{sec-fus}). }

%
%
%

%


\medskip
Our contributions are the following:
\begin{itemize}
\item We first define a sound and complete query rewriting operator for UCQs and disjunctive existential rules, which has the advantage of establishing a tight relationship between a chase step and a rewriting step
(Theorem \ref{th-sound-complete-disjunctive-rewriting}). The associated breadth-first query rewriting algorithm outputs a minimal UCQ-rewriting when one exists 
\mlmm{(Theorem \ref{th-algo-rewriting})}. 
\item We then turn our attention to the notion of \emph{finite unification sets (fus)}, which denotes sets of existential rules for which any UCQ is UCQ-rewritable, i.e., admits a finite sound and complete rewriting under the form of a UCQ. Noting that the known \emph{fus} classes for conjunctive existential rules do not seem to be generalizable to disjunctive rules, 
\mlmm{
we show that, in fact, for \emph{any} ``truly disjunctive'' nonrecursive rule, there is a CQ that is not UCQ-rewritable}
(Theorem \ref{th-non-fus}). This leads to question the relevance of \emph{fus} for disjunctive rules and to consider the problem of whether a specific UCQ is UCQ-rewritable.  
\item Finally, considering (disjunctive) mappings, we show that the problem of determining whether a given UCQ on the target schema admits a UCQ-rewriting on the source schema is undecidable (Theorem \ref{th-map-rewrite-undec}).
\end{itemize}
Based on these results, we conclude with a number of open problems. 



\section{Preliminaries}


\paragraph{Generalities.}
We consider logical vocabularies of the form $\mathcal V = (\mathcal P, \mathcal C)$, where $\mathcal P$ is a finite set of predicates and  $\mathcal C$ is a (possibly infinite) set of constants.  A \emph{term} on $\mathcal V$ is a constant from $\mathcal C$ or a variable. An
\emph{atom} on $\mathcal V$ has the form $p(\vect{t})$ where $p \in \mathcal P $ is a predicate of arity $n$ and $\vect{t}$ is a tuple of terms on $\mathcal V$ with $|\vect{t}|=n$. 
An atom with predicate $p$ is also called a $p$-atom.
Given a formula or set of formulas $S$, we denote by $\vars{S}$, $\consts{S}$ and $\terms{S}$ its sets of variables, constants and terms, respectively. 
We will often see a tuple $\vect{x}$ of pairwise distinct variables as a set.  
We denote by $\models$ and $\equiv$ classical logical  entailment and equivalence, respectively. 
Given two sets of atoms $S_1$ and $S_2$,
a \emph{homomorphism} $h$ from $S_1$ to $S_2$ is a substitution of $\vars{S_1}$
by $\terms{S_2}$ such that $h(S_1) \subseteq S_2$ (we say that $S_1$ \emph{maps} to 
 $S_2$ by $h$).
 It is well-known that, when we see $S_1$ and $S_2$ as existentially closed conjunctions of atoms, $S_2 \models S_1$ iff $S_1$ maps to $S_2$. 
 
 A \emph{safe copy} of an atom set $S$ is obtained from $S$ by a bijective renaming of its variables with \emph{fresh} variables (i.e., that do not occur elsewhere in the context of the computation).

 
 \paragraph{Knowledge base.}
A \emph{set of facts} $\fb$ is a possibly infinite set of atoms, logically seen as an existentially closed conjunction. When this set is finite we call it a \emph{fact base}.
%
%
A \emph{disjunctive existential rule} $R$ (or simply rule hereafter)  is a closed formula of  the form  
  $$\forall\vect{x}\forall\vect{y} ~(~B[\vect{x},\vect{y}] \rightarrow \bigvee\limits^n_{i=1} \exists \vect{z_i} H_i[\vect{x{_i}}, \vect{z_i}]~)$$
 where $n \geq 1$, $B$ and the $H_i$ are non-empty finite conjunctions of atoms with $\vars{B} = \vect{x} \cup \vect{y}$ and $\vars{H_i} = \vect{x{_i}} \cup \vect{z_i}$, 
$\vect{x}=\bigcup\limits_{i=1}^n \vect{x_i}$ and $\vect{x}, \vect{y}$ and the $\vect{z_i}$ are pairwise disjoint;
 $B$ is the \emph{body} of $R$, also denoted by $\body{R}$, and $\{H_1, \ldots, H_n\}$ is the \emph{head} of $R$,  also denoted by $\head{R}$. We also denote by $\headi{i}{R}$ the i-th disjunct $H_i$ of the head of $R$.
 The set $\vect x$ is the \emph{frontier} of $R$ and is denoted by $\fr{R}$. Its elements are called frontier variables. 
The set $\vect{z_i}$ is the set of \emph{existential variables} of $H_i$, also denoted by $\exist{H_i}$, and the union of all the $\exist{H_i}$ is the set of existential variables of $R$, also denoted by $\exist{R}$. 
  Note that constants may occur anywhere. For brevity, we often denote by $B \rightarrow H_1 \lor \ldots \lor H_n$ a rule with body $B$ and head $\{H_1, \ldots, H_n\}$. A rule $R$ is \emph{conjunctive} if $n=1$.
A (disjunctive) rule $R$ is (disjunctive) \emph{datalog} if $\exist{R} = \emptyset$.   
  
A (disjunctive) \emph{knowledge base} (KB) is a pair $(\fb,\mathcal R)$, where $\fb$ is a fact base 
and $\mathcal R$ is a finite set of (disjunctive) existential rules. We assume w.l.o.g. that distinct rules in $\mathcal R$ have disjoint sets of variables. In examples, we may reuse variables for simplicity. 



\paragraph{Disjunctive chase.} 

 A rule $R = B \rightarrow H_1 \lor \ldots \lor H_n$ is \emph{applicable} on a fact base $\fb$ if there is a homomorphism $h$ from $\body{R}$ to $\fb$.
The pair $(R, h)$ is called a \emph{trigger} on $\fb$.
The application of $(R,h)$ 
to $\fb$ is denoted by $\alpha_\vee(\fb, R, h)$; it produces a set of $n$ fact bases, each obtained by adding to $\fb$ a set of atoms obtained from $\headi{i}{R}$ by replacing each frontier variable $x$ by $h(x)$ and each existential variable by a fresh variable. We denote by $h^{\mathtt{safe}_i}$ the extension of $h$ that safely renames $\exist{\headi{i}{R}}$ by fresh variables. Then:

$$\alpha_\vee(\fb, R, h) = \{\fb \cup h^{\mathtt{safe}_i}(\headi{i}{R}) ~|~ 1 \leq i \leq n\}$$
The disjunctive chase procedure iteratively applies triggers towards a fixpoint. 
This procedure is often seen as the construction of a tree, see in particular \cite{BOURHIS16,CARRAL17}.  



\begin{definition}[Derivation tree]
 
A \emph{derivation tree} $\mathcal{T}$ of a KB  $(\fb,\mathcal R)$
is a (possibly infinite) rooted labeled tree $(V, E, \lambda)$, where $V$ is the set of vertices, $E$ the set of edges, and $\lambda$ a vertex labeling function inductively defined as follows:
\begin{itemize}
\item $\lambda(r) = \fb$ for the root $r$ of $\mathcal T$;
\item For each vertex $v$ with children $\{v_1, ..., v_n\}$, there is a trigger $(R, h)$ on $\lambda(v)$ with $R = B \rightarrow H_1 \lor \ldots \lor H_n  \in \mathcal{R}$ and the restriction of $\lambda$ to the domain $\{v_1, ..., v_n\}$ is a bijection to $\alpha_\vee(\lambda(v),R,h)$. 
\end{itemize}
\end{definition}
Note that we do not impose any criterion of trigger applicability, as we do not aim at studying a particular chase strategy.
%
A \emph{branch} $\gamma$ of a rooted tree is a maximal path from the root; we denote by $\textit{nodes}(\gamma)$ its set of vertices. Given a derivation tree $\mathcal T$, we denote by $\Gamma(\mathcal{T})$ the set of all its branches. A trigger $(R,h)$ on $\fb$ is \emph{satisfied} (by $\fb$)  if there is an extension $h'$ of $h$ with $h'(\headi{i}{R}) \subseteq \fb$ for some $i$. A derivation tree $(V, E, \lambda)$ is \emph{fair} if, 
 for each branch $\gamma$ and each vertex $v \in \textit{nodes}(\gamma)$, 
any trigger on $\lambda(v)$ is satisfied in a $\lambda(v')$ with $v'\in \textit{nodes}(\gamma)$. 
Finally, a \emph{chase tree} is a fair derivation tree.

\begin{definition}[Disjunctive chase result] 
The \emph{result of a disjunctive chase} of $\fb$ by $\mathcal R$ is $\textit{chase}(F, \mathcal{R}) = \{ \bigcup\limits_{v \in \textit{nodes}(\gamma)} \lambda(v) ~|~ \gamma \in \Gamma(\mathcal{T})\}$ where $\mathcal{T}$ is a chase tree and $\lambda$ its labeling function.
\end{definition}

From a logical viewpoint, the chase result is a \emph{disjunction} of existentially closed conjunctions of atoms.
Neither the chase tree nor the chase result are unique, however all the results entail the same queries (see next Theorem \ref{th-link-models-disjunctive-chase}). 
Although the degree of each  
vertex in a chase tree is bounded by the maximal number of disjuncts in a rule head, the tree may have infinite branches, and an infinite number of them. When the chase tree is finite, the result of the chase is the (finite) set of fact bases associated with its leaves.


It is sometimes convenient to consider a linearization of a finite derivation tree, which we call a derivation. 
A \emph{derivation}  of $(\{\fb\},\mathcal R)$ is a finite sequence of sets of fact bases and triggers $\mathcal{D} = (\mathcal F_0 = \{\fb\}) \xrightarrow{t_1}{} \mathcal F_1  \xrightarrow{t_2}{} \dots \xrightarrow{t_k}{} \mathcal F_k $ where $t_i = (R,h)$ is a trigger of $R \in \mathcal R$ on an $\fb_{j} \in \mathcal F_{i-1}$ and $\mathcal F_{i} = (\mathcal F_{i-1} \setminus \{\fb_{j}\} ) \cup \alpha_\lor(\fb_{j},R,h)$, for all $1\leq i \leq k$.  
To each finite derivation tree can be assigned a derivation obtained from any total ordering of the trigger applications associated with the inner 
 vertices in the tree, in a compatible way with the parent-child partial order.
%
%
%
%
 When $\mathcal R$ is a set of conjunctive rules, a derivation tree is a path and the  $\mathcal F_i$ in a derivation are singletons; then, a derivation can be seen as a sequence of fact bases (instead of sets of fact bases). 
%
%
%


\paragraph{Query Answering.}


\mlmm{
A \emph{conjunctive query} (CQ) $\qCQ$ takes the form
$\exists \vect{y}~\phi[\vect{x},\vect{y}]$, where  $\vect{x}$ and $\vect{y}$ are disjoint tuples of variables, and $\phi$ is a finite conjunction of atoms with  $\vars{\phi} = \vect{x} \cup \vect{y}$. The variables in $\vect x$ are called \emph{answer variables}. 
}
 A \emph{Boolean CQ} has no answer variables. In a \emph{full} CQ, all variables are answer variables. 
An \emph{atomic} CQ has a single atom.  
A (Boolean) \emph{union of conjunctive queries} (UCQ) is a disjunction 
 of (Boolean) CQs with \mlmm{the same tuple of answer variables $\vect{x}$}. 
For clarity, we denote a UCQ by $\qUCQ$  and a CQ by $\qCQ$. 
 A set of  facts $\fb$ \emph{answers positively} to a Boolean CQ $\qCQ$  if  $\fb \models \qCQ$. 
More generally, a tuple of constants $\vect{c}$  is an \emph{answer} to a CQ $\qCQ$ on $\fb$ if there is a substitution $s$ such that $s(\vect{x}) = \vect{c}$ and 
$\fb \models s(\qCQ)$. 
This extends to a UCQ $\qUCQ$ and a set of sets of facts $\mathcal F$: 
a tuple of constants $\vect{c}$ is an answer to $\qUCQ$ on $\mathcal F$ if for every $\fb_i \in \mathcal F$, there is a CQ $\qCQ_j \in \qUCQ$ such that $\vect{c}$ is an answer to $\qCQ_j$ on $\fb_i$. 







W.l.o.g. we focus in this paper on Boolean queries, to avoid technicalities related to answer variables. 
Hence, in the following, by UCQ and CQ we refer to \emph{Boolean} queries, unless otherwise specified. We will often see a CQ as a set of atoms, and a UCQ as a set of atoms sets.



The following theorem states that the disjunctive chase provides a sound and complete procedure to decide whether a UCQ is entailed by a disjunctive KB. 

\begin{theorem}[from \cite{BOURHIS16}]\label{th-link-models-disjunctive-chase}
Let $\mathcal Q$ be a (Boolean) UCQ and $(\fb, \mathcal{R})$ be a disjunctive KB. Then $\fb, \mathcal{R} \models \mathcal Q$ iff $\textit{chase}(\fb, \mathcal{R}) \models \mathcal Q$ {$($i.e., $F_i \models  \mathcal Q$ for all  $F_i \in \textit{chase}(\fb, \mathcal{R}))$.}
\end{theorem}

\begin{example}[Colorability] \label{ex-color}
Let $\fb$ be a fact base on predicates $v$ (vertex) and $e$ (edge) describing a graph $G$. Let $R = v(x) \rightarrow g(x) \lor r(x)$ (``Every vertex has color green or red'').  Then, $\textit{chase}(\fb, \{R\})$ yields all ways of coloring each vertex. 
Let the UCQ $\qUCQ = \{ \qCQ_1, \qCQ_2\}$ with $\qCQ_1 = \{g(u),e(u,w),g(w)\}$ and $\qCQ_2 =  \{r(u),e(u,w),r(w)\}$. The KB $(\fb, \{R\})$ answers positively to $\qUCQ$ iff $G$ is not 2-colorable. 
\end{example}

Given UCQs $\qUCQ_1$ and $\qUCQ_2$,  we say that $\qUCQ_1$ is \emph{more specific} than $\qUCQ_2$ if $\qUCQ_1 \models \qUCQ_2$. 
Note that $\mathcal Q_1\models \mathcal Q_2$ iff for all $Q_1\in \mathcal Q_1$, there is  $Q_2\in \mathcal Q_2$ such that $Q_1 \models Q_2$ (i.e., $Q_2$ maps to $Q_1$ by homomorphism).
 A CQ $\qCQ$ is \emph{minimal} if it has no strict subset $\qCQ' \subsetneq \qCQ$ such that $\qCQ' \equiv \qCQ$ (i.e.,  $\qCQ' \models \qCQ$). 
A UCQ $\qUCQ$ is \emph{minimal} if it has no strict subset $\qUCQ' \subsetneq \qUCQ$ such that $\qUCQ \equiv \qUCQ'$ (whether each CQ in the UCQ is itself minimal is not relevant for our results). 
A \emph{cover} of a UCQ $\qUCQ$ is a minimal subset $\qUCQ' \subseteq \qUCQ$ such that  $\qUCQ \equiv \qUCQ'$. 
It is known that, given two equivalent UCQs $\qUCQ_1$ and $\qUCQ_2$, there is a bijection from any cover of $\qUCQ_1$ to any cover of $\qUCQ_2$ that maps each CQ in $\qUCQ_1$ to an equivalent CQ in $\qUCQ_2$ (see, e.g., \cite{SWJ15}).

\paragraph{Mappings.}
Given two disjoint sets of predicates $\mathcal S$ and $\mathcal T$, respectively called the \emph{source} and the \emph{target} predicates, a \emph{source-to-target} (or $\mathcal S$-to-$\mathcal T$) rule $R$ is such that $\body{R}$ uses predicates in $\mathcal S$ and  $\head{R}$ uses predicates in $\mathcal T$. 
A (disjunctive) mapping $\mathcal M$ on $(\mathcal S, \mathcal T)$ is a finite set of $\mathcal S$-to-$\mathcal T$ (disjunctive) rules. 
In this setting, a fact base (or database instance) is expressed on  $\mathcal S$ and a query on $\mathcal T$. 
Note that the chase of a fact base with a mapping is always finite. 
 
 
 \vspace*{-0.1cm}
\paragraph{UCQ rewritability. }
In the following, by \emph{rewriting} of a UCQ $\qUCQ$ with a set of rules $\mathcal R$, we mean a possibly infinite set of CQs $\qUCQ'$, 
such that for all fact base $\fb$, if $\fb \models \qUCQ'$ then $\fb, \mathcal R\models \qUCQ$ (in other words, a rewriting is by definition \emph{sound}). 
A rewriting  $\qUCQ'$ of $\qUCQ$ with $\mathcal R$ is \emph{complete} if for all 
fact base $\fb$, if $\fb, \mathcal R \models \qUCQ$ then $\fb \models \qUCQ'$.
A finite complete rewriting is called a \emph{UCQ-rewriting}.  A pair $(\qUCQ,\mathcal R)$ is called \emph{UCQ-rewritable} if it admits a UCQ-rewriting. The set $\mathcal R$ itself is called \emph{UCQ-rewritable} if for any UCQ $\qUCQ$, the pair $(\qUCQ,\mathcal R)$ is UCQ-rewritable.  In the framework of conjunctive existential rules, a UCQ-rewritable set is also called a \emph{finite unification set (fus)} \cite{AIJ11}. We shall extend this term to disjunctive rules. 

\begin{example}[Transitivity] \label{ex:transitivity}
Let $R = p(x,y) \land p(y,z) \rightarrow p(x,z)$.  The (Boolean) CQ $\qCQ_1= \{ p(a,b) \}$, where $a$ and $b$ are constants, has no UCQ-rewriting with $\{R\}$, while the (Boolean)  CQ $\qCQ_2 = \{ p(u,v) \}$ has one, which is $\{\qCQ_2 \}$. 
Indeed, any complete rewriting of $\qCQ_1$ is infinite as it contains all the ``paths'' of $p$-atoms from $a$ to $b$, which are pairwise incomparable by homomorphism. 
In contrast, the atom $p(u,v)$ maps by homomorphism to any path of $p$-atoms.  
\end{example}

Finally, we recall some fundamental notions on rewriting with \emph{conjunctive} existential rules. We will rely on these to define rewriting with disjunctive rules. 
 
 \vspace*{-0.1cm}
\paragraph{Query rewriting with conjunctive existential rules}  In the setting of conjunctive existential rules, query rewriting can be performed using  \emph{piece-unifiers}; these are a generalization of classical
unifiers that take care of existential variables in rule heads by unifying sets of atoms instead of single atoms \cite{DBLP:conf/iccs/SalvatM96,blms:09}.
 In short, a piece-unifier unifies a subset $Q'$ of a 
 CQ $\qCQ$ and a subset $H'$ of a rule head, such that existential variables from $H'$ are unified only with variables of $Q'$ that do not occur in $\qCQ \setminus Q'$. 
  Next, we call \emph{separating variables} of $Q'$ (w.r.t. $\qCQ$) the variables of $Q'$ that also occur in $\qCQ \setminus Q'$. 
It is convenient to represent a unifier as a partition of a set of terms rather than a substitution. Hence, we say that a partition $P$ of a set of terms is \emph{admissible} if no class of $P$ contains two constants; 
\mlmm{ we associate a substitution $u$ with an admissible partition $P_u$ by selecting} 
one term in each class with priority given to constants:  for each class $C$ in $P_u$, let $t_i$ be the selected term, then for every $t_j \in C$, we set $u(t_j) = t_i$. 

\begin{definition}[Piece-unifier]\label{def-piece-unifier}\footnote{In non-Boolean queries, answer variables have to be treated as separating variables.}
  Let $\qCQ$ be a CQ and $R = B \rightarrow H$ be a conjunctive existential rule such that $\vars{\qCQ} \cap \vars{B \cup H} = \emptyset$. A \emph{piece-unifier} of $\qCQ$ with $R$   
  is a triple $\mu = (Q',H', P_u)$ with $Q' \neq \emptyset$, $Q' \subseteq Q$, 
  $H' \subseteq H$, and 
  $P_u$ is an admissible partition on $\terms{Q'} \cup \terms{H'}$ such that:
  \begin{enumerate}
   \item $u(Q') = u(H')$, with $u$ a substitution associated with $P_u$; 
  \item  If a class $C \in P_u$ contains an existential variable (from $H'$), then the other terms in $C$ are \mlmm{non-separating} variables from $Q'$. 
  \end{enumerate}
\end{definition}

%

Let $\mu = (\qCQ',H', P_u)$ be a piece-unifier of $\qCQ$ with $R: B \rightarrow H$ and $u$ a substitution associated with $P_u$.  
The application of $\mu$ produces the following CQ: 
$$\beta(\qCQ,R,\mu)  = u(B) \cup u(\qCQ \setminus \qCQ')$$


\begin{example}[Piece-Unifier] Let $R = p(x,y) \rightarrow \exists z~p_1(x,z) \land p_2(y,z)$ and $\qCQ_1 = \{p_1(u,v), s(v)\}$. There is no piece-unifier of $\qCQ_1$ with $R$ since $v$ is a separating variable of $\qCQ'_1 = \{p_1(u,v)\}$, hence cannot be unified with $z$. Let $\qCQ_2 = \{ p_1(u,v), s(u)\}$: now, there is a piece-unifier of $\qCQ_2$ with $R$, namely $\mu_2=(\{p_1(u,v)\},\{p_1(x,z)\},P_{u_2})$ with $P_{u_2}=\{\{x,u\},\{y\},\{z,v\}\}$. Taking the substitution $u_2=\{u\mapsto x, v \mapsto z\}$, we obtain 
$\beta(Q_2,R,\mu_2)=\{p(x,y), s(x)\}$.  
Finally, let $\qCQ_3 = \{ p_1(u, v), p_2(u, w), p_1(t,v), s(t)\}$, and $\qCQ'_3 = \qCQ_3 \setminus \{s(t)\}$. 
The triple 
$\mu_3=(\qCQ'_3,\head{R},P_{u_3})$ with 
 $P_{u_3} = \{\{x,y,t,u\}, \{z,v,w\}\}$ 
is a piece-unifier of $\qCQ_3$ with $R$. If we select $x$ and $z$ in  $P_{u_3}$, $\beta(Q_3,R,\mu_3)=\{p(x,x),s(x)\}$.
\end{example}
%


A \emph{piece-rewriting} of  a UCQ $\qUCQ$ with a (conjunctive) rule set $\mathcal R$ is a UCQ $\qUCQ_k$ obtained by a finite sequence of piece-unifier applications, i.e.,  
$(\qUCQ_0=\qUCQ), \ldots, \qUCQ_k$ ($k \geq 0$) such that, for all $0 < i \leq k$, there is a piece-unifier $\mu$ of $\qCQ \in \qUCQ_{i-1}$ with $R \in \mathcal R$ such that $\qUCQ_i = \qUCQ_{i-1} \cup \{\beta(\qCQ,R,\mu)\}$. 

As stated below, piece-unifiers provide a sound and complete query rewriting procedure: 

\begin{theorem}[from \cite{AIJ11}] For any (conjunctive) KB $(\fb, \mathcal R)$ and 
UCQ $\qUCQ$, there is a derivation of $(\fb, \mathcal R)$ leading to an $\fb_i$ such that $\fb_i \models \qUCQ$ iff there is a piece-rewriting $\qUCQ_j$ of $\qUCQ$ with $\mathcal R$ such that $\fb \models \qUCQ_j$. 
\end{theorem}

It follows that, when a pair $(\qUCQ,\mathcal R)$ is UCQ-rewritable, a UCQ-rewriting can be obtained as a piece-rewriting. 
 Let us point out that a conjunctive mapping is always UCQ-rewritable (or \emph{fus}). Indeed, since it is made of $\mathcal S$-to-$\mathcal T$ rules, the application of a piece-unifier of a CQ $Q$ produces a CQ with strictly fewer atoms on $\mathcal T$ than $Q$. Also, CQs that contain predicates on $\mathcal T$ are useless in a rewriting.

%
%
%
%

%
%
%
%
%
%


\section{Query Rewriting with Disjunctive Rules}\label{sec-algorithm_ucq_rewriting}

%
%


Our generalization of query rewriting to disjunctive rules relies on a simple idea: a query $\qUCQ$ can be rewritten with a rule $R = B \rightarrow H_1 \lor \dots \lor H_n$ if \emph{each} $H_i$ contributes to partially answer $\qUCQ$.
Therefore, a unification step consists of unifying \emph{each} $H_i$ (using a piece-unifier) with a safe copy $\qCQ_i$ of a CQ from $\qUCQ$ ; safe copies ensure that the CQs involved in the unification have pairwise disjoint sets of variables. Note that several  safe copies of the same CQ from $\qUCQ$ can be involved. This yields a new CQ made of $\body{R}$ and the remaining parts of the unified CQs, according to some aggregation of the piece-unifiers. 
We need a few auxiliary notions to specify this aggregation. 
Let $\mathcal{P}$ be a set of partitions (not necessarily of the same set). The \emph{join} of $\mathcal{P}$, denoted by $\text{join}(\mathcal{P})$, is the partition obtained from $\mathcal{P}$ by making the union of the partitions in $\mathcal{P}$, then merging all non-disjoint classes until fixed point. 
\mlmm{E.g., given $\mathcal{P}$ composed of partitions $\{\{x,u\}, \{y,v\}, \{z,w\}\}$ and  $\{\{x, y,a\}, \{z',t\}\}$, we obtain  
$\textit{join}(\mathcal{P}) = \{\{x, u, y, v, a\}, \{z, w\}, \{z',t\}\}$.
}
We say that a set of partitions associated with piece-unifiers is \emph{admissible} 
 if its join is an admissible partition (i.e., it does not contain a class with two constants). 

\begin{definition}[Disjunctive Piece-Unifier and One-step Piece-Rewriting]\label{def-DPU}
Let a rule $R = B \rightarrow H_1 \lor \dots \lor H_n$ and a UCQ $\qUCQ$.
A \emph{disjunctive piece-unifier} $\mu_\lor$ of $\qUCQ$ with $R$ is a set $\{\mu_1, \ldots, \mu_n\}$ such that:
\begin{itemize}
\item for $1 \leq i \leq n$, 
$\mu_i = (\qCQ_i', H_i', P_{u_i})$ is a (conjunctive) piece-unifier of $\qCQ_i$, a safe copy of a CQ from $\qUCQ$, with the (conjunctive) rule $B \rightarrow H_i$;
\item and $\mathcal{P}_{u_\lor}= \{P_{u_1}, \ldots,P_{u_n}\}$  is admissible. 
\end{itemize}
Given a substitution $u_{\lor}$  associated with $\text{join}(\mathcal{P}_{u_\lor})$, the application of $\mu_\lor$ produces 
 the CQ 
$$\beta_\lor(\qUCQ, R, \mu_\lor) = u_{\lor}(B) \cup \bigcup\limits_{1 \leq i \leq n} u_{_\lor}(\qCQ_i\setminus \qCQ_i')$$
The \emph{one-step piece-rewriting} 
of $\qUCQ$ w.r.t. $\mu_\lor$ is  
$$\qUCQ \cup \{ \beta_\lor(\qUCQ, R, \mu_\lor) \}$$
\end{definition}


\begin{example}
Let $R = p(x,y) \rightarrow \exists z_1~r(x,z_1)~\lor~\exists z_2~r(y,z_2)$ and the UCQ $\qUCQ = \{\qCQ\}$ with $\qCQ = \{s(u), r(u,v)\}$. Let $\qCQ_1 = \{s(u_1), r(u_1,v_1)\}$ and $\qCQ_2 = \{s(u_2),r(u_2,v_2)\}$ be two safe copies of $\qCQ$, and let 
$\mu_\lor = \{\mu_1, \mu_2\}$ with 
$\mu_1 = (\{r(u_1,v_1)\},\, \{r(x,z_1)\},\, $ $\{\{u_1,x\},\,  \{v_1,z_1\}\})$ and $\mu_2 = (\{r(u_2,v_2)\},\,  \{r(y,z_2)\},\, $  $\{\{u_2,y\},\,  \{v_2,z_2\}\})$. 
Assume we 
give priority to variables from $R$, i.e., we take the substitution $u_\lor = \{ u_1 \mapsto x,$ $v_1 \mapsto z_1,~ u_2 \mapsto y,~ v_2 \mapsto z_2 \}$. Then $\beta_\lor(\qUCQ, R, \mu_\lor) = \{p(x,y),s(x) ,s(y)\}$. 
\end{example}

\begin{definition}[Piece-Rewriting] Given a disjunctive rule set $\mathcal R$, a  UCQ $\qUCQ'$ is a \emph{piece-rewriting} 
(or simply \emph{rewriting} when clear from the context)  of a UCQ $\qUCQ$ with $\mathcal R$ if there is a finite sequence (called \emph{rewriting sequence}) $\qUCQ = \qUCQ_0, \qUCQ_1,\dots,\qUCQ_k = \qUCQ'$ $(k \geq 0)$, such that for all $0 < i \leq k$, there is  a disjunctive piece-unifier $\mu_\lor$ 
of $\qUCQ_{i-1}$ with $R \in \mathcal{R}$ such that 
$\qUCQ_i$ is the one-step rewriting of $\qUCQ_{i-1}$ w.r.t. $\mu_\lor$. 
\end{definition}
The following lemmas highlight fundamental properties of $\alpha_\lor$ and $\beta_\lor$. 

\begin{lemma}[Preservation of entailment by $\alpha_\lor$ and $\beta_\lor$]\label{lemma-entail}
Let $R$ be a disjunctive rule. 
\begin{enumerate}
\item For any fact bases $F_1$ and $F_2$ such that $\fb_2\models \fb_1$: if there is a trigger $(R,\homo_1)$ on $\fb_1$ then there is a trigger $(R,\homo_2)$ on $\fb_2$ such that $\alpha_\lor(\fb_2,R,\homo_2)\models \alpha_\lor(\fb_1,R,\homo_1)$.
\item For any UCQs $\qUCQ_1$ and  $\qUCQ_2$ such that  $\qUCQ_2\models \qUCQ_1$: if there is a (disjunctive) piece-unifier $\mu_2$ of $\qUCQ_2$ with $R$ then either $\beta_\lor(\qUCQ_2,R,\mu_2) \models \qUCQ_1$, or
there is a (disjunctive) piece-unifier $\mu_1$ of $\qUCQ_1$ with $R$ such that $\beta_\lor(\qUCQ_2,R,\mu_2) \models  
\beta_\lor(\qUCQ_1,R,\mu_1)$. 
\end{enumerate}
\end{lemma}

The second lemma clarifies the tight relationship between $\alpha_\lor$ and $\beta_\lor$ (we recall that fact bases and CQs have the same logical form; this is also true of finite sets of fact bases and UCQs). 

\begin{lemma}[Composition of $\alpha_\lor$ and $\beta_\lor$]\label{lemma-compo}
Let $R$ be a disjunctive rule. 
\begin{enumerate}
\item For any fact base  $\fb$: if there is a trigger $(R,\homo)$ on $\fb$ then there is a (disjunctive) piece-unifier $\mu$ of $\alpha_\lor(\fb,R,\homo)$ with $R$ such that $\fb\models \beta_\lor(\alpha_\lor(\fb,R,\homo),R,\mu)$.
\item For any UCQ $\qUCQ$: if there is a piece-unifier $\mu$ of $\qUCQ$ with $R$ then there is a trigger $(R,\homo)$ on $\beta_\lor(\qUCQ,R,\mu)$ such that $\alpha_\lor(\beta_\lor(\qUCQ,R,\mu),R,\homo)\models\qUCQ$.
\end{enumerate}
\end{lemma}

These two lemmas are keys to establish the soundness and completeness of piece-rewriting, as stated next.

\begin{theorem}[Soundness and completeness of piece-rewriting]
\label{th-sound-complete-disjunctive-rewriting}
Let  $\mathcal R$ be a set of disjunctive rules and $\qUCQ$ be a UCQ. Then, for any fact base $\fb$, holds $\fb, \mathcal{R} \models \qUCQ$ iff there is a piece-rewriting $\qUCQ'$  of $\qUCQ$ such that $\fb\models \qUCQ'$. 
\end{theorem}

\begin{figure}[t]
\begin{tikzpicture}
\node (fb) at (0,0) {$\fb$};
\node (alphafb) at (2, 0) {\color{darkgray}$\alpha_\lor(\fb, R, \homo)$};
\node (Q) at (2,2) {$\qUCQ$};
\node (betaQ) at (0,2) {$\beta_\lor(\qUCQ, R, \mu_\lor)$};
\graph  {
  (fb) ->[dashed, red, "$\alpha_\lor$"] (alphafb) ->[dashed, "$\models$"] (Q),
  (Q) ->[blue, sloped, "$\beta_\lor$"] (betaQ),
  (fb) ->[below, "$\models$"] (betaQ),
};
\end{tikzpicture}
\begin{tikzpicture}
\node (fb) at (0,0) {$\fb$};
\node (alphafb) at (2, 0) {$\alpha_\lor(\fb, R, \homo)$};
\node (Q) at (2,2) {$\qUCQ$};
\node (betaQ) at (0,2) {\color{darkgray}$\beta_\lor(\qUCQ, R, \mu_\lor)$};
\node (models) at (0,1) {$~~~~\models$};
\graph  {
  (fb) ->[red, "$\alpha_\lor$"] (alphafb) ->["$\models$"] (Q),
  (fb) ->[dashed, sloped, bend left=20] (Q),
  (Q) ->[dashed, blue, sloped, "$\beta_\lor$"] (betaQ),
  (fb) ->[dashed]  (betaQ)
  
};
\end{tikzpicture}
\caption{Correspondences between $\beta_\lor$ (in blue) and $\alpha_\lor$ (in red)}
\label{fig-alphabeta}
\end{figure}
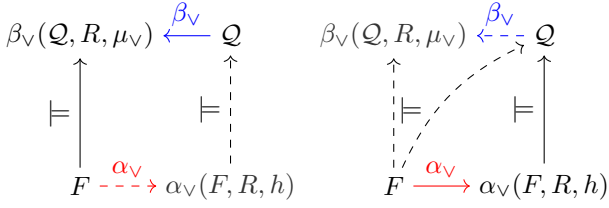

\begin{proof}(Sketch) We show that there is a derivation of $(\{F\}, \mathcal R)$ leading to an $\mathcal F_i$ such that $\mathcal F_i \models \qUCQ$ iff there is a rewriting $\qUCQ_j$ of $\qUCQ$ with $\mathcal R$ such that $\fb\models \qUCQ_j$ (with moreover $j \leq i$).  This equivalence relies on the following two lemmas, which are corollaries of previous Lemmas \ref{lemma-entail} and \ref{lemma-compo}.
Given any Boolean UCQ $\qUCQ$, disjunctive rule $R$ and fact base  $\fb$, the following holds (see Figure \ref{fig-alphabeta}):
\begin{itemize}
\item (Backward-forward Lemma) 
For any disjunctive piece-unifier $\mu_\lor$ of $\qUCQ$ with $R$, if $\fb \models \beta_\lor(\qUCQ, R, \mu_\lor)$ then there is a trigger $(R,\homo)$ on $\fb$ such that $\alpha_\lor(\fb, R, \homo) \models \qUCQ$; 
\item (Forward-backward Lemma) 
For any trigger $(R,\homo)$ on $\fb$,  if $\alpha_\lor(\fb, R, \homo) \models \qUCQ$ then either $\fb \models \qUCQ$ or there is a disjunctive piece-unifier $\mu_\lor$ of $\qUCQ$ with $R$, such that $\fb \models \beta_\lor(\qUCQ, R, \mu_\lor)$.
\end{itemize}

The ($\Rightarrow$) direction of the theorem is proved by induction on the length $k$ of a derivation 
from $\{\fb\}$ to $\mathcal F_k$ such that 
$\mathcal F_k \models \qUCQ$, using forward-backward Lemma (which itself follows from Lemma \ref{lemma-compo} (Point 1) and Lemma \ref{lemma-entail} (Point 2)). The ($\Leftarrow$) direction is proved by induction on the length $k$ of a rewriting sequence from $\qUCQ$ to $\qUCQ_k$ such that $\fb\models \qUCQ_k$, using backward-forward Lemma (which itself follows from Lemma \ref{lemma-compo} (Point 2) and Lemma \ref{lemma-entail} (Point 1)). 
\end{proof}

To actually compute a UCQ-rewriting of $\qUCQ$ when one exists,
it is convenient to proceed in a breadth-first manner, i.e., extend $\qUCQ$ at each step with all the CQs that can be generated with (new) disjunctive piece-unifiers.  More specifically, we inductively define the following operator $W$, which takes as input a UCQ $\qUCQ$ and a disjunctive rule set $\mathcal R$, and returns a possibly infinite set of CQs:
\begin{itemize}
\item $W_0(\qUCQ, \mathcal R) = \qUCQ$
\item For $i > 0$, $W_i(\qUCQ, \mathcal R) = W_{i-1}(\qUCQ, \mathcal R) \cup$ $\{ \beta_\lor(W_{i-1}(\qUCQ, \mathcal R), R, \mu_\lor) | \mu_\lor \text{ piece-unifier with } R \in \mathcal R  \}$ 
\item Finally, $W(\qUCQ, \mathcal R) =   \bigcup\limits_{i \in \mathbb{N}} W_i(\qUCQ, \mathcal R)$.
\end{itemize}


\begin{proposition}[Properties of $W$]\label{prop-UCQ-rewriting}
For any UCQ $\qUCQ$ and disjunctive rule set $\mathcal R$, the following holds:
\begin{enumerate}
\item $W(\qUCQ, \mathcal R)$ is a complete rewriting of $(\qUCQ, \mathcal R)$. 
\item If $(\qUCQ, \mathcal R)$ admits a UCQ-rewriting $\mathcal Q'$, then there is $i \geq 0$ such that $\mathcal Q' \equiv W_i(\qUCQ, \mathcal R)$. 
\end{enumerate}
\end{proposition}

\begin{proof} (1) Each $W_i(\qUCQ, \mathcal R)$ is a piece-rewriting of $\qUCQ$ with $\mathcal R$  and, for any piece-rewriting $\qUCQ'$ of $\qUCQ$ with $\mathcal R$, there is $i$ such that $\qUCQ' \subseteq W_i(\qUCQ, \mathcal R)$. Hence, the union of all the $W_i(\qUCQ, \mathcal R)$ is a complete rewriting of $\qUCQ$. (2) If $(\qUCQ, \mathcal R)$ admits a UCQ-rewriting $\mathcal Q'$, then by Theorem \ref{th-sound-complete-disjunctive-rewriting} it admits a complete piece-rewriting $\qUCQ''$, and both are necessarily equivalent. Then, $\qUCQ'' \subseteq W_i(\qUCQ, \mathcal R)$ for some $i$ and, since $\qUCQ''$ is complete,  $\qUCQ'' \equiv W_i(\qUCQ, \mathcal R)$. 
\end{proof}

We propose a query rewriting algorithm (see Algorithm \ref{algo-rewriting}) that mimics the computation of $W(\qUCQ, \mathcal R)$, while including two optimizations at each step $i > 0$. First, it only considers \emph{new} disjunctive piece-unifiers, i.e., those that involve at least one CQ generated at step $i-1$.
Second, it removes redundant CQs in the rewriting under construction, by the computation of a cover.   
 More specifically, $\qUCQ^{\star}$ denotes the rewriting under construction 
and $\qUCQ_{new}$  the set of CQs generated at a given step.
The function $\cover{}$ (Lines 1 and 6) returns a cover of the given set.  The function $\generate{}$ (Line 5) takes as input the current rewriting $\qUCQ^{\star}$, its subset $\qUCQ_{prev}$ of CQs generated at the previous step,  as well as $\mathcal R$, 
and returns the set of generated CQs, i.e., all the $\beta_\lor(\qUCQ^{\star}, R, \mu_\lor)$ where  $\mu_\lor$ is a new disjunctive piece-unifier.  
This yields the set $\qUCQ_{new}$. 
To compute a cover of $\qUCQ^{\star} \cup \qUCQ_{new}$, priority is given to $\qUCQ^{\star}$ in case of query equivalence, for termination reasons. The function $\removeMoreSpecific{}$ takes as input two sets of CQs and returns the first set minus its queries more specific than a query of the second set.  The computation of a cover of $\qUCQ^{\star} \cup \qUCQ_{new}$ is decomposed into three steps (Lines 6-8): compute a cover of $\qUCQ_{new}$; remove from $\qUCQ_{new}$ the queries more specific than a query from $\qUCQ^{\star}$; and remove from  $\qUCQ^{\star}$ the queries more specific than a query from $\qUCQ_{new}$. Then, $\qUCQ_{new}$ is added to $\qUCQ^{\star}$ (Line 9). 
We remind that a query may have rewritings of unbounded size but still a UCQ-rewriting (see Example \ref{ex:transitivity}), hence the role of the cover computation is not only to remove redundancies but also to ensure that the algorithm halts when a UCQ-rewriting has been found. 

\begin{algorithm}
\KwData{UBCQ $\qUCQ$ and set of disjunctive rules $\mathcal R$}
\KwResult{A sound and complete rewriting of $\qUCQ$}
$\qUCQ_{new} \leftarrow \cover(\qUCQ)$;  \emph{// new CQs}\\
$\qUCQ^{\star} \leftarrow \qUCQ_{new}$; \emph{// result}\\
\While{$\qUCQ_{new} \neq \emptyset$}
{
$\qUCQ_{prev} \leftarrow \qUCQ_{new}$ \emph{// CQs from the preceding step}\\
$\qUCQ_{new} \leftarrow \generate{}(\qUCQ^{\star},\qUCQ_{prev},\mathcal R)$; \emph{// new CQs}\\
$\qUCQ_{new}\leftarrow \cover(\qUCQ_{new})$\\
$\qUCQ_{new}\leftarrow \removeMoreSpecific(\qUCQ_{new},\qUCQ^{\star})$\\
$\qUCQ^{\star}\leftarrow \removeMoreSpecific(\qUCQ^{\star},\qUCQ_{new})$\\
$\qUCQ^{\star} \leftarrow\qUCQ^{\star} \cup \qUCQ_{new}$
}
\textbf{return} $\qUCQ^{\star}$
\caption{{\sc Breadth-First Rewriting}} \label{algo-rewriting}
\end{algorithm}

The correctness of the algorithm is  based  on the soundness and completeness of the $W$ operator, however 
attention should be paid to the potential impact of query removal on completeness (Lines 6 to 8). 
Indeed, when a CQ $\qCQ_2$ is removed because it is more specific than another CQ $ \qCQ_1$, 
we have to ensure that any CQ that could be generated using $\qCQ_2$ is more specific than another CQ already present in the curent rewriting, or than a CQ that can be generated using $\qCQ_1$. 
\mlmm{Fortunately, this property is ensured by Lemma \ref{lemma-entail} (Point 2), considering $\mathcal Q^\star$ and $\mathcal Q_{new}$ at the end of Line 5, then taking $\mathcal Q_2$ = $\mathcal Q^\star\cup \mathcal Q_{new}$ 
and $\qUCQ_1 = \qUCQ_2 \setminus \{\qCQ_2\}$.}

\begin{theorem}\label{th-algo-rewriting}

Algorithm \ref{algo-rewriting} computes a sound and complete rewriting. Moreover, it halts and outputs a minimal rewriting when $(\qUCQ, \mathcal R)$ is UCQ-rewritable.  
\end{theorem}

\begin{proof} By induction on the number of iterations of the while loop, we prove the following invariant of the algorithm, using 
Lemma \ref{lemma-entail} (Point 2): after step $i$, $\qUCQ^{\star}$ is equivalent to $W_i(\qUCQ, \mathcal R)$. Then, soundness and completeness follow from Proposition \ref{prop-UCQ-rewriting}. 
Line 7 ensures that $\qUCQ_{new}$ becomes empty when $\qUCQ^{\star}$ is a complete rewriting. Since a cover of $\qUCQ^{\star}$ is computed at each step, the output set is of minimal size. 
\end{proof}

\paragraph{Further remarks on completeness.}  When it comes to practical implementations, one may find simpler to rely on (conjunctive) piece-unifiers that unify the smallest possible subsets of a CQ. Such piece-unifiers are called \emph{single-piece} \cite{SWJ15}. In the specific case of datalog, a single-piece unifier unifies a single atom of a CQ with a rule head. Piece-rewriting restricted to single-piece unifiers is complete for conjunctive rules \cite{SWJ15}, but it is no longer so with disjunctive rules. 
This occurs already in the case of disjunctive datalog, as illustrated next. 


%

\begin{example}\label{ex-color2} Consider again the colorability example (Ex. \ref{ex-color}) with $R = v(x) \rightarrow g(x) \lor r(x)$ and
$\qUCQ = \{ \qCQ_1, \qCQ_2\}$ with $\qCQ_1 = \{g(u),e(u,w),g(w)\}$ and $\qCQ_2 =  \{r(u),e(u,w),r(w)\}$.
With single-piece unifiers we obtain CQs that have the shape of ``chains'' with a $g$-atom or an $r$-atom at each extremity. 
However, there are also rewritings without any occurrence of $g$ nor $r$, and the only way of obtaining them is to unify two query atoms together.
For instance, the CQ $\{v(u), e(u,u) \}$ is obtained by unifying, on the one hand both $g$-atoms of a safe copy of $\qCQ_1$ with $g(x)$, and on the other hand
both $r$-atoms of a safe copy of $\qCQ_2$ with $r(x)$. 
More generally, using such piece-unifiers, one can produce all the CQs that describe the odd-length cycles in the graph. Note that these CQs are incomparable with the CQs generated with single-piece unifiers. This example also shows that a UCQ may have no UCQ-rewriting although each of its CQs has one (which is here the CQ itself).  
\end{example}

\vspace*{- 0.1cm}
{\paragraph{Related work.} 
To the best of our knowledge, \cite{ALFONSO21} is the only previous work proposing a UCQ rewriting technique for general disjunctive existential rules. 
This technique is based on a restricted form of first-order resolution, where  at each step a CQ is unified with a disjunct of a rule head (using a conjunctive piece-unifier), which produces a new disjunctive rule with fewer disjunctions; when the unified rule is conjunctive, (the negation of) a CQ is produced.  In comparison, the main advantages of our proposal are the following: \emph{(1)} a rewriting step directly produces a CQ and not a rule,\emph{ (2) } intermediate rules, which may not lead to a CQ, are avoided, and \emph{(3)} there is a direct correspondence between a chase step and a rewriting step, which makes it easier to study the properties of query rewriting, especially as the rule set is not updated. 
}


%



\section{What are \emph{fus} Disjunctive Rules?}\label{sec-fus}

We now address the question of identifying classes of disjunctive rules that are UCQ-rewritable. By extension of the term coined for conjunctive existential rules, we also call them \emph{fus}. 
To the best of our knowledge, the only \emph{fus} class of disjunctive rules mentioned in the literature \cite{ALFONSO21} is actually a slight extension of \emph{fus} conjunctive rules: this class consists of disjunctive rules with an empty frontier and it is shown that such rules can be safely added to a set of \emph{fus} conjunctive rules. 
As a matter of fact, known \emph{fus} classes of conjunctive rules do not seem to be extensible to the disjunctive case. And worse, the straightforward extension of syntactic criteria that underlie \emph{fus} in the conjunctive case seems to easily lead to undecidability of query answering, as shown for example in \cite{MORAK21} for the syntactic restriction called stickiness \cite{DBLP:journals/pvldb/CaliGP10}.

At first glance, one may expect \emph{nonrecursive} disjunctive rule sets to be \emph{fus}, as it happens for conjunctive rules. However, it is not the case, 
as shown by the next example: a CQ (on unary predicates) may have no UCQ-rewriting even with a \emph{single non-recursive body-atomic} (disjunctive) \emph{datalog} rule. 

\begin{example} \label{ex-not-fus}
 Let the rule $R = p(x,y) \rightarrow t_1(x) \lor t_2(y)$ and the BCQ $\qCQ = \{t_1(u), t_2(u)\}$. 
Then the pair $(\{\qCQ\},\{R\})$ has no UCQ-rewriting. 
Indeed, a complete rewriting contains all the CQs of the following shape for any $n \in \mathbb{N}$:

$$t_2(u_0) \land \left(\bigwedge\limits_{i = 1}^n p(u_{i-1}, u_{i}) \right) \land t_1(u_n) $$


All these queries are pairwise incomparable w.r.t. homomorphism. Let us detail the first rewriting step. To unify $\{\qCQ\}$ with $R$, we have to make two safe copies of  $\qCQ$, let $\qCQ_1$ and $\qCQ_2$, which are respectively unified with $t_1(x)$ and $t_2(y)$. This produces the CQ $\{t_2(x), p(x,y), t_1(y)\}$, isomorphic to $\{t_2(u_0), p(u_0,u_1), t_1(u_1)\}$.  If we switch the unified atoms of $\head{R}$, we obtain an isomorphic CQ.  
All subsequent rewriting steps lead to longer paths of $p$-atoms. 
\end{example}

\mlmm{
A similar observation follows from \cite{DBLP:conf/kr/GerasimovaKKPZ20}, which focuses on a specific disjunctive rule of the form $A(x) \rightarrow T(x) \lor F(x)$, called a covering axiom and denoted by $cov_A$; their complexity results imply that the singleton set $\{ cov_A \}$ is not \emph{fus},\footnote{That paper studies syntactic conditions on ontology-mediated CQs of the form $(Q,cov_A)$ that determine the data complexity of query answering and the rewritability in some target query language. In particular, it is shown that if a (connected) CQ $Q$ has no term $x$ with both atoms $T(x)$ and $F(x)$  and contains at least one $F$-atom and one $T$-atom then answering $(Q, cov_A)$ is L-hard for data complexity. Since answering a UCQ-rewritable ontology-mediated query is in $AC^0$ for data complexity, and $AC^0 \subset L$, it follows that no $cov_A$ is fus. }  
 which can be checked for instance by considering the query 
$Q = \{ T(u), p(u,v), F(v) \}$. 
}
%
%
%
%
%
%
%
%
%

\mlmm{Next, we show that such observations can be generalized to almost \emph{any} source-to-target disjunctive rule.}
Evidently, we have to exclude disjunctive rules that are equivalent to a conjunctive rule, as classes of \emph{fus} conjunctive rules are known. We also exclude \emph{disconnected} rules, i.e., rules $R$ such that  $\body{R} \cup  \head{R}$ is not a connected set of atoms (where connectivity is defined in the obvious way based on shared variables). Note that a rule with a head $H_i$ that has an empty frontier is disconnected, as well as a rule whose body has a connected component with an empty frontier.  However,  a rule with a disconnected body may not be disconnected, since head atoms may connect several connected components of the body 
(e.g., a ``product'' rule like $b_1(x) \land b_2(y) \rightarrow t_1(x) \lor t_2(y) \lor p(x,y)$ is not disconnected). 

\begin{example}[\emph{Fus} disconnected rule] Let the disconnected rule $R= b(x) \rightarrow t_1(x) \lor \exists z~t_2(z)$. $R$ is not equivalent to a conjunctive rule. Let us check that it is \emph{fus}. Given any UCQ $\qUCQ$, let $\qUCQ_2$ be the subset of $\qUCQ$ that contains all the CQs that can be unified with $\exists z~t_2(z)$. Any  $Q \in \qUCQ_2$ necessarily contains a disconnected component of the form $\exists u~t_2(u)$. Moreover, it is useless to unify  $Q$ with $t_1(x)$: in such case, let $Q_2$ be the CQ unified with $\exists z~t_2(z)$, then the obtained rewriting is more specific than $Q_2$. Hence, we can ignore all the produced CQs that contain a connected component of the form $\exists u~t_2(u)$. Rewriting $\qUCQ$ with $\{R\}$ amounts to rewriting $\qUCQ \setminus  \qUCQ_2$ with the conjunctive rule set 
$\mathcal{R} = \{b(x) \land (\qCQ_2 \setminus \{\exists u~t_2(u)\}) \rightarrow t_1(x) ~|~ \qCQ_2 \in \qUCQ_2\}$, which belongs to the \emph{fus} class called \emph{domain restricted} \cite{AIJ11}. 
%
%
\end{example}






%

In the next theorem, we restrict the head of the rule to a disjunction of two atom sets, to keep the proof simple. 

\begin{theorem} \label{th-non-fus} Let $R = B \rightarrow H_1 \lor H_2$ be a source-to-target rule that is not disconnected nor equivalent to a conjunctive rule.
Then, there is a CQ $\qCQ$ such that  $(\{\qCQ\},\{R\})$ is not UCQ-rewritable. 
\end{theorem}

\begin{proof}(Sketch)
Let $R = B[\vect{x_1}, \vect{x_2}, \vect{y}] \rightarrow \exists  \vect{z_1}~H_1[ \vect{x_1},  \vect{z_1}] \lor \exists  \vect{z_2}~H_2[ \vect{x_2},  \vect{z_2})]$, where:
\begin{itemize}
\item $\fr{R} =  \vect{x_1} \cup  \vect{x_2}$; $ \vect{x_1}$ and $ \vect{x_2}$ may share variables;
\item $ \vect{x_i} \neq \emptyset$ ($i = 1,2$) since $R$ is not disconnected. 
\end{itemize}
We build the following (Boolean) CQ:


$$\qCQ = \{ H^s_1[ \vect{v_1},  \vect{w_1}], p( \vect{v_1},  \vect{v_2}), H^s_2[ \vect{v_2},  \vect{w_2}] \}$$

where each $H^s_i[ \vect{v_i},  \vect{w_i}]$ is a safe copy of $H_i[ \vect{x_i},  \vect{z_i}]$ and $p$ is a fresh predicate. 
Note that, since $R$ is connected, both $H_1$ and $H_2$ have a frontier variable, and frontier variables being renamed in each $H^s_i$, the arity of $p$ is at least $2$. 
In $p( \vect{v_1},  \vect{v_2})$ the order on the variables is important: a fixed order is chosen on $ \vect{x_i}$ (hence, $ \vect{v_i}$) and the tuple $ \vect{v_1}$ comes before the tuple $ \vect{v_2}$. Hence, $p( \vect{v_1}, \vect{v_2})$ can be seen as ``directed'' from  $\vect{v_1}$ to $\vect{v_2}$. 
We then proceed in two steps.
\begin{enumerate}
\item We show that we can produce an infinite set $\qUCQ$ whose element CQs are pairwise incomparable by homomorphism. 
Let $\qCQ_0 = \qCQ$. 
At each step $i \geq 1$, $\qCQ_i$ is produced from a safe copy of $\qCQ$ unified with $H_1$ and a safe copy of $\qCQ_{i-1}$ unified with $H_2$. The piece-unifiers unify $H_1^s$ (resp. $H_2^s$) in $\qCQ$ (resp. $\qCQ_{i-1}$) according to the isomorphism from $H_1^s$ (resp. $H_2^s$) to $H_1$ (resp. $H_2$). 
Any CQ $\qCQ_k$ in $\qUCQ$ is connected and follows the ``pattern'' $H^s_2.p.(B.p)^k.H^s_1$, where occurrences of $p$-atoms all have the same direction; hence, two ``adjacent'' $p$-atoms, i.e., that share variables with the same copy $B_i$ of a $B$, cannot be mapped one onto the other (by a homomorphism that maps $B_i$ to itself).  
\item We show that no CQ $\qCQ'$ that can be produced by piece-rewriting maps by homomorphism to a CQ from $\qUCQ$, except by isomorphism. 
When there is no (conjunctive) piece-unifier that unifies $H_1[ \vect{v_1},  \vect{w_1}]$ in $\qCQ$ with $H_2[ \vect{x_2},  \vect{z_2}]$ (the same holds if we exchange $H_1$ and $H_2$), all the produced $\qCQ'$ are more specific than (including isomorphic to) CQs from $\qUCQ$. 
Otherwise, assume that a CQ $Q'$ is produced by unifying $H_1[ \vect{v_1},  \vect{w_1}]$ with $H_2[ \vect{x_2},  \vect{z_2}]$. If $Q'$ can be mapped by homomorphism to a $Q_n \in \mathcal Q$, the arguments of any $p$-atom in $Q'$ must be pairwise distinct variables. We show that it leads to have $R$ equivalent to the conjunctive rule $B \rightarrow H_i$ (with $i =1$ or $i=2$), which contradicts the hypothesis on $R$. 
%
\end{enumerate}
It follows that $\qUCQ$ is a subset of any sound and complete rewriting of $\{\qCQ\}$ with $\{R\}$, hence the pair $(\{\qCQ\},\{R\})$ does not admit a UCQ-rewriting.
\end{proof}

\mlmm{One interest of the above proof is to provide a general construction that applies to any rule (fulfilling the conditions of the theorem). Also, the proof can be generalized to a rule head with $k$ disjuncts, taking $\qCQ$ containing a safe copy of each $H_i$ plus a $p$-atom that connects these copies through their frontier variables. }



Given this result, the notion of \emph{fus} disjunctive rules does not seem to be particularly relevant. Studying the problem of deciding whether a pair $(\mathcal Q, \mathcal R)$ is UCQ-rewritable seems more interesting, although it is known to be undecidable already for (conjunctive) datalog rules.\footnote
{This follows from the undecidability of determining whether a datalog program is uniformly bounded \cite{GAIFMAN93}. Indeed, a datalog program $\mathcal R$ is uniformly bounded iff the pair $(\qCQ,\mathcal R)$ is UCQ-rewritable for any \emph{full} atomic query $\qCQ$. 
In turn, UCQ-rewritability of $(\qCQ,\mathcal R)$ can be reduced to UCQ-rewritability of $(\qCQ',\mathcal R)$ with $\qCQ'$ a Boolean CQ.
}
Again, little is known about classes of disjunctive rules and UCQs for which this problem would be decidable. 
Let us point out a few immediate cases of UCQ-rewritable pairs $(\mathcal Q, \mathcal R)$: 
\begin{itemize}
\item $\mathcal Q$ is composed of  atomic CQs and $\mathcal R$ is a set of disjunctive linear existential rules (i.e., rules with an atomic body). Indeed, only atomic CQs can be produced, and there is a finite number of them on a given set of predicates. \mlmm{This case was already noticed in \cite{BOURHIS16}. }
\item $\mathcal Q$ is composed of  atomic queries and $\mathcal R$ is a set of $\mathcal S$-to-$\mathcal T$ rules. The produced CQs are obtained from the rule bodies by specializing their frontier (i.e., merging variables and replacing them by constants occurring in   $\mathcal Q$ and rule heads). Hence, there is a finite number of them.  
\item $\mathcal Q$ is composed of variable-free CQs\footnote{If non-Boolean CQs are considered, $\mathcal Q$ can be extended to a set of full CQs.}
 and $\mathcal R$ is a set of lossless existential rules (i.e., such that all the variables in a rule body are frontier). Then, no variable is introduced by rewriting, hence the number of terms in a CQ is bounded by $|\consts{\mathcal Q} \cup \consts{\mathcal R}|$. 
%
\end{itemize}

\section{Disjunctive Mappings}
\label{sec-disjunctive-mappings}

We now consider UCQ-rewritability with (disjunctive) mappings.
Let $\mathcal S$ and $\mathcal T$ be the sets of source and target predicates, respectively, and let $\mathcal M$ be a mapping  on $(\mathcal S,\mathcal T)$. 
Given a query on $\mathcal T$, the aim is to obtain a complete rewriting w.r.t. fact bases on $\mathcal S$. Because $\mathcal S$ and $\mathcal T$ are disjoint, CQs that contain atoms on $\mathcal T$ are useless in a rewriting. Hence, we define a \emph{mapping rewriting} as a rewriting on $\mathcal S$ and use the notation \emph{$\mathcal S$-rewriting} to distinguish it from a rewriting on $\mathcal S \cup \mathcal T$. An $\mathcal S$-rewriting $\qUCQ'$ of a UCQ $\mathcal Q$ with $\mathcal M$ is \emph{complete} if, for all fact base $\fb$ on $\mathcal S$, if $\fb, \mathcal M \models \qUCQ$ then $\fb \models \qUCQ'$. A finite complete $\mathcal S$-rewriting is called a \emph{UCQ-$\mathcal S$-rewriting}.
%
%
 



\begin{example}[Colorability] We adapt Example \ref{ex-color2} to transform the rule into a mapping. Let $\mathcal S = \{v,e\}$, $\mathcal T = \{\hat{e}, g,r\}$ and $\mathcal M = \{m_1, m_2\}$, with:

$m_1= e(x,y) \rightarrow \hat{e}(x,y)$

$m_2= v(x) \rightarrow g(x) \lor r(x)$. 

Let $\qUCQ = \{ \qCQ_1, \qCQ_2\}$ with $\qCQ_1 = \{g(u),\hat{e}(u,w),g(w)\}$ and $\qCQ_2 =  \{r(u),\hat{e}(u,w),r(w)\}$.
Any complete $\mathcal S$-rewriting of $\qUCQ$ contains CQs that describe all the cycles of odd length (in other words, it defines non-2-colorability). 
All the other CQs that can be produced by piece-rewriting contain predicates $g$ and $r$, hence are discarded.  
\end{example}

Note that a query may have a UCQ-$\mathcal S$-rewriting, while it does not have any UCQ-rewriting (on $\mathcal S \cup \mathcal T$), as illustrated by the next example.


\begin{example} Let $\mathcal S= \{p\}$ and $\mathcal T = \{t_1,t_2\}$. Consider the (Boolean) CQ $\qCQ = \{t_1(u), t_2(u)\}$ and the rule $R = p(x,y) \rightarrow t_1(x) \lor t_2(y)$ from Example \ref{ex-not-fus}.
While the pair $(\{\qCQ\}, \{R\})$ has no UCQ-rewriting, it has a UCQ-$\mathcal S$-rewriting, which is empty. Indeed, all the CQs that can be obtained by piece-rewriting contain an atom on $\mathcal T$. 
\end{example}

Let \emph{disjunctive mapping rewritability} be the following problem: Given a disjunctive mapping $\mathcal M$ on $(\mathcal S,\mathcal T)$ and a UCQ $\qUCQ$ on $\mathcal T$, does $(\qUCQ, \mathcal M)$ have a UCQ-$\mathcal S$-rewriting ?


\begin{theorem}\label{th-map-rewrite-undec}
Disjunctive mapping rewritability is undecidable.
\end{theorem}

\begin{proof} (Sketch) We build a reduction from the following undecidable problem: Given a (Boolean) CQ $\qCQ$ and a set of (conjunctive) datalog rules $\mathcal R$, is the pair $(\{\qCQ\}, \mathcal R)$ UCQ-rewritable? 
W.l.o.g. we assume that rules in $\mathcal R$ have no constants (and an atomic head).  
The reduction translates each instance $(\qCQ, \mathcal R)$
defined on a set of predicates $\mathcal P$, into an instance $(\mathcal Q^{Q,\mathcal R},\mathcal M^{Q,\mathcal R})$ of the disjunctive mapping rewritability problem, defined on a pair \mlmm{of predicats sets} $(\mathcal S, \mathcal T)$ such that:
%
\begin{itemize}
\item  $\mathcal S = \mathcal P\cup\{T\}$, where $T$ is a fresh unary predicate, 
\item  $\mathcal T$ is the union of: (1) a set of predicates in bijection with $\mathcal S$, 
where $\hat{p}$ denotes the predicate obtained from $p \in \mathcal S$,
and (2) a set of fresh predicates in bijection with $\mathcal R$, where $p_{R_i}$ denotes the predicate associated with the rule $R_i$; 
the arity of each $p_{R_i}$ is $|\fr{R_i}|$. 
\end{itemize}

Given a conjunction $Q$ (on $\mathcal P$), we denote by $Q^T$ the conjunction (on $\mathcal S$) obtained from $Q$ by adding a $T$-atom on each term; 
 given a conjunction $Q$ (on $\mathcal S$), we denote by 
 $\hat Q$ the conjunction (on $\mathcal T$) obtained from $Q$ by renaming all the predicates $p$ into $\hat{p}$. 
Hence, $\widehat{Q^T}$ is obtained by performing the first operation, then the second. 
Given $\vect{x} = x_1, \ldots, x_n$, $T[\vect{x}]$ denotes the conjunction $T(x_1) \land \dots \land T(x_n)$. 
Similarly, $\hat{T}[\vect{x}] = \hat{T}(x_1) \land \dots \land \hat{T}(x_n)$.

Let 
$\qCQ$
 and $\mathcal R= \{R_1, \dots, R_n\}$, where $R_i = B_i[\vect{x_i}, \vect{y_i}] \rightarrow H_i[\vect{x_i}]$. 
The instance $(\qUCQ^{\qCQ,\mathcal R},\m^{\qCQ,\mathcal R})$ is defined as follows:
\begin{itemize}
\item $\qUCQ^{\qCQ,\mathcal R} = ~~  \{\qCQ_\qCQ\} \cup \qUCQ_\mathcal{R}$ with: \\
    \quad \quad $\qCQ_Q = \widehat{\qCQ^T}$,   \\
    \quad \quad $\mathcal {Q_R}=\{\qCQ_{R_i} = \exists \vect{x_i}, \vect{y_i} ~ \widehat {{(B_i)}^T}[\vect{x_i}, \vect{y_i}] \land p_{R_i}(\vect{x_i}) | R_i \in \mathcal R\}$    
\item $\m^{\qCQ,\mathcal R} = \m_\mathcal R \cup \m_{trans}$ with: \\
    $~~\m_\mathcal R =\{ m_{R_i} = T[\vect{x_i}] \rightarrow p_{R_i}(\vect{x_i}) \lor \hat H_i(\vect{x_i}) ~|~ R_i \in \mathcal R\}$ \\
    $~~\m_{trans} = \{p(\vect{x}) \rightarrow \hat p(\vect{x}) ~|~ p \in \mathcal S\}$
\end{itemize}

Based on the natural bijection between the CQs $Q_\mathcal P$ defined on $\mathcal P$ and the CQs \mlmm{$(Q_\mathcal P)^T$} defined on $\mathcal S$,  
we prove that 
$Q_\mathcal P$ belongs to a rewriting of $\{\qCQ\}$ with $\mathcal R$ iff \mlmm{$(Q_\mathcal P)^T$} belongs 
to a rewriting of $\qUCQ^{\qCQ,\mathcal R}$ with $\m^{\qCQ,\mathcal R}$. 
Note that set membership is up to isomorphism throughout the proof. 
More specifically, we first prove the following lemmas:
\begin{enumerate}
\item For any CQ $Q_w$ in a piece-rewriting of $\{\qCQ\}$ with $\mathcal R$, \mlmm{$(Q_w)^T$} belongs to a piece-rewriting of $\mathcal Q^{\qCQ,\mathcal R}$ with $\m^{\qCQ, \mathcal R}$. Indeed, to each $R_i$ are associated a CQ $\qCQ_{R_i}$ and a rule $m_{R_i}$ 
that allow to simulate any rewriting step performed with $R_i$, using fresh predicate $p_{R_i}$.
\item Any CQ $Q_S$ in an $\mathcal S$-rewriting of $\mathcal Q^{\qCQ,\mathcal R}$ with $\m^{\qCQ, \mathcal R}$ is of the form $Q_S =$ \mlmm{$(Q_\mathcal P)^T$}, with $Q_\mathcal P$ the subset of $Q_S$ on $\mathcal P$.
\item For any CQ of the form \mlmm{$(Q_\mathcal P)^T$}, with $Q_\mathcal P$ on $\mathcal P$, that belongs a piece-rewriting of $\mathcal Q^{\qCQ,\mathcal R}$ with $\m^{\qCQ, \mathcal R}$, $Q_\mathcal P$ belongs to a piece-rewriting of $\{\qCQ\}$ with $\mathcal R^\star$, where $\mathcal R^\star$ is the reflexive and transitive closure of $\mathcal R$ by unfolding (i.e., rule composition). Note that $\mathcal R^\star$ is logically equivalent to $\mathcal R$. 
\end{enumerate}

We rely on these lemmas to prove the following: if there is a UCQ-rewriting of $(\{Q\},\mathcal R)$ then there is a UCQ-$\mathcal S$-rewriting of $(\mathcal Q^{\qCQ,\mathcal R},\m^{\qCQ, \mathcal R})$. The proof of the opposite direction is similar.
Let $\mathcal Q$ be a UCQ-rewriting of $(\{Q\},\mathcal R)$. Then there is a piece-rewriting $\mathcal Q_i$ of $\{Q\}$ with $\mathcal R$ such that $\mathcal Q_i\equiv \mathcal Q$.
By Lemma 1, there is a piece-rewriting $\mathcal Q_j$ of $\mathcal Q^{\qCQ,\mathcal R}$ with $\m^{\qCQ, \mathcal R}$ 
that contains all the CQs of the form $(Q_w)^T$ in bijection with the $Q_w$ in $\mathcal Q_i$. By definition, 
$\mathcal Q_j$ is a finite rewriting of $(\mathcal Q^{\qCQ,\mathcal R},\m^{\qCQ, \mathcal R})$ and the subset $\mathcal Q^{\mathcal S}_j$ of $\mathcal Q_j$ that contains only the CQs on $\mathcal S$ is a finite $\mathcal S$-rewriting of $(\mathcal Q^{\qCQ,\mathcal R},\m^{\qCQ, \mathcal R})$. 
Now, assume $\mathcal Q^{\mathcal S}_j$ is not complete, i.e., 
there is a CQ 
that belongs to an $\mathcal S$-rewriting of $(\mathcal Q^{\qCQ,\mathcal R},\m^{\qCQ, \mathcal R})$ but that is not more specific than a CQ in $\mathcal Q^{\mathcal S}_j$; by Lemma 2, such CQ is of the form $(Q_\mathcal P)^T$. 
Then there is a piece-rewriting $\mathcal Q'_j$ of $\mathcal Q^{\qCQ,\mathcal R}$ with $\m^{\qCQ, \mathcal R}$ that contains a CQ 
entailed by $(Q_\mathcal P)^T$; hence such CQ is also on  $\mathcal S$, and by Lemma 2 it is of the form $(Q'_\mathcal P)^T$. 
By Lemma 3, $Q'_\mathcal P$ belongs to a piece-rewriting of $\{\qCQ\}$ with $\mathcal R^\star$. 
Since $\mathcal R^\star \equiv \mathcal R$, there is a CQ equivalent to $Q'_\mathcal P$ in some rewriting of $(\{Q\},\mathcal R)$. Since 
%
$\mathcal Q_i$ is complete, 
there is $Q_c \in \mathcal Q_i$ such that $Q'_\mathcal P\models Q_c$. Hence, $(Q'_\mathcal P)^{T}\models (Q_c)^T$, so $(Q_\mathcal P)^{T}\models (Q_c)^T$; by Lemma 1, $(Q_c)^T \in \mathcal Q_j$, hence $(Q_c)^T \in \mathcal Q^{\mathcal S}_j$, which contradicts the fact that  $(Q_\mathcal P)^T$ is not more specific than a CQ in $\mathcal Q^{\mathcal S}_j$. 
\end{proof}

\mlmm{\section{Perspectives}}

\mlmm{In conclusion,} UCQ rewriting with disjunctive existential rules appears to be extremely challenging. The main classes that ensure termination for conjunctive rules fail to be generalized. As suggested by \mlmm{previous work in \cite{DBLP:conf/kr/GerasimovaKKPZ20} and our} Theorem \ref{th-non-fus}, the \emph{fus} notion applied to disjunctive rules does not seem to add much w.r.t. \emph{fus} conjunctive rules.  However, it might be more relevant in the context of mappings (when it becomes UCQ-$\mathcal S$-rewritability), which still has to be studied. Beside, a number of interesting issues remain open, in relationship with the finite rewritability of a pair $(\mathcal Q, \mathcal R)$. We list here some of them:
\begin{enumerate}
\item Clarify the boundary between decidability and undecidability for the problem of determining whether a pair $(\mathcal Q, \mathcal R)$ is UCQ-rewritable, according to specific classes of rules (and queries).  In particular, UCQ-rewritability is decidable for guarded conjunctive rules and some of their generalizations  \cite{DBLP:conf/ijcai/BarceloBLP18}, does this extend to the disjunctive case?
\item We have shown that the UCQ-$\mathcal S$-rewritability of a pair $(\mathcal Q, \mathcal M)$ is undecidable (Theorem \ref{th-map-rewrite-undec}). Is it still the case for a pair  $(\{Q\}, \mathcal M)$ where $Q$ is a CQ? 
\item Our undecidability proof for UCQ-$\mathcal S$-rewritability (Theorem \ref{th-map-rewrite-undec}) exploits the fact that rewritings are restricted to predicates in $\mathcal S$. If we consider instead UCQ-rewritings with source-to-target rules, we know that the problem can only be simpler, as there is an easy reduction from UCQ-rewritability with $\mathcal S$-to-$\mathcal T$-rules to UCQ-$\mathcal S$-rewritability with mappings (one simply has to add a mapping rule per target predicate to give it an existence at the source level). Is the UCQ-rewritability of a pair $(\mathcal Q, \mathcal R)$ decidable when $\mathcal R$ is a set of  $\mathcal S$-to-$\mathcal T$ rules? 
\item Design an algorithm that, given a pair $(\mathcal Q, \mathcal M)$, 
outputs a UCQ-$\mathcal S$-rewriting for this pair when one exists. 
\end{enumerate}

\mlmm{
\section*{Acknowledgements} This work is partly supported by the ANR project CQFD (ANR-18-CE23-0003).
} 
%
%
%
%
%
%
%
%

\bibliographystyle{kr}
\bibliography{bibliography}

\onecolumn
\newpage
\appendix
\section*{Appendix}

\bigskip


\section{Proofs of Section \ref{sec-algorithm_ucq_rewriting}}


\mic{In these proofs, we reuse some notations and results from \cite{AIJ11} and \cite{SWJ15}.

Let $h : X \longrightarrow T$ and $h':X'\longrightarrow T'$ be two substitutions such that, $\forall x \in X \cap X',h(v) = h'(v)$. Then we note $h+h':X\cup X'\longrightarrow T \cup T'$ the substitution defined by: if $x \in X, (h+h')(x) = h(x)$, otherwise $(h+h')(x) = h'(x)$.
}

\begin{proposition}[\mic{was Prop. 23 in} \cite{AIJ11}]\label{prop-concat-homomorphisms} Let $\fb$ be a fact base, $\qCQ$ be a CQ, $\vect{x} \subseteq \vars{\qCQ}$, $\{\qCQ_1, \dots, \qCQ_k\}$ be a partition of the atoms of $\qCQ$ such that 
 $\vars{\qCQ_i}\cap \vars{\qCQ_j} \subseteq \vect{x}$ for all $\qCQ_i$ and $\qCQ_j$ with $i \neq j$, 
and $\homo_1, \dots, \homo_k$ homomorphisms from $\qCQ_i$ to $\fb$ such that, $\forall t \in \vect{x}, \forall 1 \leq i \leq j \leq k, \homo_i(t) = \homo_j(t)$; then the substitution $\homo_1 + \dots + \homo_k$ is a homomorphism from $\qCQ$ to $\fb$.
\end{proposition}


Given a partition $P$ on a set of terms, we denote by $P[t]$ the class of $P$ containing the term $t$.

\begin{definition}[Partition induced by a substitution] A partition $P$ on terms $\mathcal{T}$ induced by a substitution $s$ is such that for every $t, t' \in \mathcal{T}$, if $s(t) = s(t')$ then $t' \in P[t]$ \mic{(\textit{i.e.} $P[t]=P[t']$)} and $P$ is the thinnest partition with this property. \mic{Let $C$ be a class of $P$, we call \emph{selected element} of $C$, which we denote $t_C$, the unique element of C such that $s(t_C)=t_C$.} 
\end{definition}

The three next propositions are immediate.

\begin{proposition}\label{prop-specialization-set-fact-bases-homomorphism}
Let $\mathcal F$ be a set of \mic{set of facts} 
and $\mathcal Q$ be a UCQ: $\mathcal F\models \mathcal Q$ iff for each $\fb\in \mathcal F$, there exists a $Q\in \mathcal Q$ such that $Q$ maps to $\fb$.
\end{proposition}

\begin{proposition}\label{prop-sub-adm} A partition induced by a substitution is admissible.
\end{proposition}


\begin{proposition}\label{prop-sub-unif} Let $\fb$ and $\fb'$ be two fact bases and $s$ a substitution from $\fb$ to $\fb'$ such that $s(\fb)=\fb'$. Then, any substitution $u_s$ associated with $P_s$, the partition induced by $s$, on the terms of $\fb$ and $\fb'$, is such that $u_s(\fb)=u_s(\fb')$.
\end{proposition}



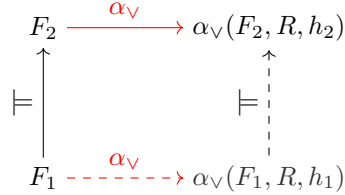
\begin{figure}[ht]
\begin{center}
\begin{tikzpicture}
\node (fb) at (0,0) {$\fb_1$};
\node (alphafb) at (3, 0) {\color{darkgray}$\alpha_\lor(\fb_1, R, \homo_1)$};
\node (fb') at (0,2) {$\fb_2$};
\node (alphafb') at (3,2) {$\alpha_\lor(\fb_2, R, \homo_2)$};
\graph  {
  (fb) ->[dashed, red, "$\alpha_\lor$"] (alphafb) ->[dashed, "$\models$"] (alphafb'),
  (fb') ->[red, "$\alpha_\lor$"] (alphafb'),
  (fb) ->[below, "$\models$"] (fb'),
};
\end{tikzpicture}
\end{center}
\caption{
\mic{Preservation of entailment by $\alpha_\lor$} (Prop. \ref{prop-alpha-specialization})\label{fig-alpha-specialization}}
\end{figure}

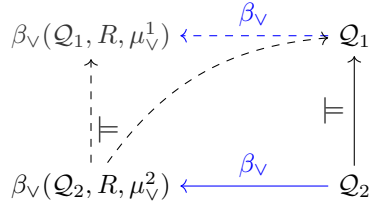
\begin{figure}[ht]
\begin{center}
\begin{tikzpicture}
\node (Q1) at (3.5,2) {$\qUCQ_1$};
\node (betaQ1) at (0,2) {\color{darkgray}$\beta_\lor(\qUCQ_1, R, \mu_\lor^1)$};
\node (Q2) at (3.5,0) {$\qUCQ_2$};
\node (betaQ2) at (0,0) {$\beta_\lor(\qUCQ_2, R, \mu_\lor^2)$};
\node (models) at (0,0.75) {$~~~~\models$};
\graph  {
  (Q1) ->[dashed, sloped, blue, "$\beta_\lor$"] (betaQ1),
  (betaQ2) ->[dashed, sloped, bend left=25] (Q1),
  (Q2) ->[blue, sloped, "$\beta_\lor$"] (betaQ2),
  (Q2) ->["$\models$"]  (Q1),
  (betaQ2) ->[dashed] (betaQ1)
};
\end{tikzpicture}
\end{center}
\caption{
\mic{Preservation of entailment by $\beta_\lor$} (Prop. \ref{prop-beta-specialization})\label{fig-beta-specialization}}
\end{figure}

The following propositions \ref{prop-alpha-specialization} and \ref{prop-beta-specialization} correspond to \textbf{Lemma \ref{lemma-entail}} (Point 1 and Point 2, respectively) in the paper. \mic{Figures \ref{fig-alpha-specialization} and \ref{fig-beta-specialization} depict these propositions.}


\begin{proposition}\label{prop-alpha-specialization}
Let $\fb_1, \fb_2$ be two fact bases such that $\fb_1 \models \fb_2$ and a disjunctive rule $R$ such that there exists a trigger $(R, \homo_2)$ on $\fb_2$. Then, there exists a trigger $(R,\homo_1)$ on $\fb_1$ such that $\alpha_\lor(\fb_1, R, \homo_1) \models \alpha_\lor(\fb_2, R, \homo_2)$. 
\end{proposition}

\begin{proof} 
Let $R = B \rightarrow H_1 \lor \dots \lor H_n$.
Since $\fb_1 \models \fb_2$, we have a homomorphism $\homo$ from $\fb_2$ to $\fb_1$.
Moreover, $(R,\homo_2)$ being a trigger on $\fb_2$, taking $\homo_1 = \homo \circ \homo_2$, we have $(R,\homo_1)$ is a trigger on $\fb_1$
and $\alpha_\lor(\fb_2, R, \homo_2) = \{\fb^i_2 = \fb_2 \cup \homo_2^{safe_{i \cdot 2}}(H_i) ~|~ 1 \leq i \leq n\}$ and $\alpha_\lor(\fb_1, R, \homo_1) = \{\fb^i_1 = \fb_1 \cup (\homo \circ \homo_2)^{safe_{i \cdot 1}}(H_i) ~|~ 1 \leq i \leq n\}$.
Let us build a homomorphism $\homo^i$ from $\fb^i_2$ to $\fb^i_1$, for  $1 \leq i \leq n$.
For each $i$, we first consider the homomorphism $\homo_{H_i}$ from $\homo_2^{safe_{i \cdot 2}}(H_i)$ to $(\homo \circ h_\alpha)^{safe_{i \cdot 1}}(H_i)$, defined as follows:

\vspace{0.2cm}

\emph{
\hspace{-0.4cm}$\forall t \in \vars{\homo_2^{safe_{i \cdot 2}}(H_i)}$:
\vspace{-0.1cm}
\begin{itemize}
    \item if $t \in \homo_2(\fr{R})$, then $\homo_{H_i}(t) = \homo(t)$;
    \item otherwise, $\homo_{H_i}(t) = ~.^{safe_{i \cdot 1}}((.^{safe_{i \cdot 2}})^{-1}(t))$.
\end{itemize}
}

$\homo$ and $\homo_{H_i}$ satisfy the conditions of Proposition \ref{prop-concat-homomorphisms} (with $\vect{x}=\homo_{H_i}(\fr{R})$). As a consequence, 
 $h^i = \homo + \homo_{H_i}$ is a homomorphism from $\fb^i_2$ to $\fb^i_1$.
 Thus, $\alpha_\lor(\fb_1, R, \homo_1) \models \alpha_\lor(\fb_2, R, \homo_2)$.
\end{proof}

\begin{proposition} \label{prop-beta-specialization}
Let $\qUCQ_1$ and  $\qUCQ_2$ be UCQs such that $\qUCQ_2 \models \qUCQ_1$, and let $R$ be a disjunctive rule. Then, for 
any disjunctive piece-unifier $\mu_\lor^2$ of $\qUCQ_2$ with $R$:
\begin{enumerate}
\item either $\beta_\lor(\qUCQ_2, R, \mu_\lor^2) \models \qUCQ_1$;
\item or, there is 
a piece-unifier $\mu_\lor^1$ of  $\qUCQ_1$ with $R$ such that $\beta_\lor(\qUCQ_2, R, \mu_\lor^2) \models \beta_\lor(\qUCQ_1, R, \mu_\lor^1)$.
\end{enumerate}
\end{proposition}

\begin{proof}
\mic{Let $R = B \rightarrow H_1 \lor \dots \lor H_n$.
Let 
$\qCQ^1_2, \dots, \qCQ^n_2$ be the safe copies of CQs in $\qUCQ_2$ of which subsets $\qCQ^{1'}_2, \dots, \qCQ^{n'}_2$ are unified with, respectively, $H_1', \dots, H_n'$, subsets of respectively $H_1, \dots, H_n$,  to define $\mu_\lor^2 = \{(\qCQ_2^{1'},H_1',P_{u_1}^2), \dots, (\qCQ_2^{n'},H_n',P_{u_n}^2)\}$ the disjunctive piece-unifier of $\qUCQ_2$ with $R$. Let $P_{u_\lor^2} = join(\{P_{u_1}^2,\ldots, P_{u_n}^2\})$ and let $u_\lor^2$ be the substitution associated with $P_{u_\lor^2}$. Let $h_1,\ldots, h_n$ be the homomorphisms associated with each $\qCQ^{i}_2$, that map a $\qCQ_1^i$ in $\qUCQ_1$ to $\qCQ^{i}_2$ (note that since each $\qCQ_2^i$ is a safe copy of a CQ in $\qUCQ_2$ then there exists a CQ $\qCQ_1^i$ in $\qUCQ_1$ that maps on it).}
We consider two cases:

\begin{itemize}
    \item Either, \mic{one of the $\qCQ^i_1$ maps by $\homo_i$ to the non-rewritten part of $\qCQ^i_2$, so this $\qCQ^i_1$ maps to the CQ added to the $\qUCQ_2$ by the one-step piece-rewriting, \textit{i.e.}} there exists $1 \leq i \leq n$ and $\qCQ^i_1 \in \qUCQ_1$ such that $\homo_i(\qCQ^i_1) \subseteq (\qCQ_2^i \setminus \qCQ_2^{i'})$, then $u_\lor^2 \circ \homo_i$ is a homomorphism from  $\qCQ^i_1$ to $u_\lor^2(\qCQ^i_2 \setminus \qCQ^{i'}_2) \subseteq \beta_\lor(\qUCQ_2,R,\mu_\lor^2)$. Thus $\beta_\lor(\qUCQ_2,R,\mu_\lor^2) \models \qCQ^i_1 \models \qUCQ_1$.

    \item Otherwise, for each $1 \leq i \leq n$, 
\mic{we now consider that $\qCQ^i_1$ is a safe copy of the CQ in $\qUCQ_1$ that maps to $\qCQ^i_2$ and $\homo_i$ is the homomorphism (extended by considering this safe renaming) from $\qCQ^i_1$ to $\qCQ^i_2$.
Let $\qCQ^{i'}_1$ be the maximal subset of $\qCQ^i_1$ that maps to $\qCQ_2^{i'}$ by $\homo_i$, \textit{i.e.}} $\qCQ^{i'}_1 \subseteq \qCQ^i_1$, $\homo_i(\qCQ^{i'}_1) \subseteq \qCQ_2^{i'}$ and $\homo_i(\qCQ_1^i \setminus \qCQ_1^{i'}) \cap \qCQ_2^{i'} = \emptyset$.
\mic{Let $H_i''$ be the maximal subset of $H_i'$ that is unified by $u^2_\lor$ with the subset $h_i(Q_1^{i'})$ of $Q_2^{i'}$, \textit{i.e.}} $H_i'' \subseteq H_i'$, $u_\lor^2(H_i'') = u_\lor^2(\homo_i(\qCQ_1^{i'}))$ \mic{and $u_\lor^2(H_i'\setminus H_i'')\cap u_\lor^2(\homo_i(\qCQ_1^{i'}))=\emptyset$}.
Let $P^1_{u_i}$ the partition induced by $u_\lor^2 \circ \homo_i$ on $\terms{H_i'' \cup \qCQ_1^{i'}}$. By construction, $\mu_i^1 = (\qCQ_1^{i'}, H_i'', P_{u_i}^1)$ is thus a piece-unifier between $\qCQ_1^i$ and $H_i$.
   Since for each $1 \leq i < j \leq n$, $\qCQ_1^{i'}$ and $\qCQ_1^{j'}$ does not share any variable, then we can define $\homo = \homo_1 + \dots + \homo_n$.
We have that $u_\lor^2 \circ \homo$ is a homomorphism from $\qCQ_1^{1'} \land \dots \land \qCQ_n^{1'}$ to $u_\lor^2(H_1'' \land \dots \land H_n'')$. Let $P_{u_\lor^1}$ be the partition induced by $u_\lor^2 \circ \homo$ on $\terms{\qCQ_1^{1'} \land \dots \land \qCQ_n^{1'}} \cup \terms{H_1'' \land \dots \land H_n''}$: it is admissible since it is built from a substitution (Proposition \ref{prop-sub-adm}). Moreover, we have $P_{u_\lor^1} = join(P_{u_1}, \dots, P_{u_n})$ and thus, $\mu_\lor^1$ is a disjunctive unifier of $\qUCQ_1$ with $R$.

We now prove that $\beta_\lor(\qUCQ_2,R,\mu_\lor^2) \models \beta_\lor(\qUCQ_1,R,\mu_\lor^1)$.
We build a substitution $s$ from the selected elements of the classes in $P_{u_\lor^1}$ which are variables, to the selected elements of the classes in $P_{u_\lor^2}$ as follows: for any class $C \in P_{u_\lor^1}$, 
\mic{if $t_C$ is a variable of a $H_i''$, then $s(t_C) = u_\lor^2(t_C)$, otherwise $s(t_C) = u_\lor^2(\homo(t))$ ($t$ occurs in a $\qCQ_i^{1'}$). Note that for any term $t$ in $P_{u_\lor^1}$, we have $s(u_\lor^1(t)) = u_\lor^2(\homo(t))$.     }
    We build now a substitution $\homo'$ from $\vars{\beta_\lor(\qUCQ_1, R, \mu_\lor^1)}$ to $\terms{\beta_\lor(\qUCQ_2, R, \mu_\lor^2)}$ by considering three cases according to the part of $\beta_\lor(\qUCQ_1, R, \mu_\lor^1)$ in which the variables occurs (in a $\qCQ_i^{1}$ but not in $\qCQ_i^{1'}$, in $\body{R}$ but not in $H_i''$, or in the remaining part corresponding to the images of $\vars{\qCQ_1^{i'}} \cap \vars{\qCQ_1^i}$ by $u_\lor^1$):

    \begin{itemize}
        \item if $x \in \vars{\qCQ_1^{i}} \setminus \vars{\qCQ_1^{i'}}$, $\homo'(x) = \homo(x)$;
        \item  if $x \in \vars{\body{R}} \setminus \vars{\bigcup_{i=1}^n H_i''}$, $\homo'(x) = u_\lor^2(x)$;
        \item if $x \in u_\lor^1(\bigcup_{i=1}^n (\vars{\qCQ_1^{i'}} \cap \vars{\qCQ_1^i}))$ (or alternatively $x \in u_\lor^1(\fr{R} \cap \vars{\bigcup_{i=1}^n H_i''})$), $\homo'(x) = s(x)$.
     \end{itemize}
     
        We conclude by showing that $\homo'$ is a homomorphism from $\beta_\lor(\qUCQ_1, R, \mu_\lor^1) = u_\lor^1(\body{R}) \cup \bigcup_{i=1}^n u_\lor^1(\qCQ_1^{i} \setminus \qCQ_1^{i'})$ to $\beta_\lor(\qUCQ_2, R, \mu_\lor^2) = u_\lor^2(\body{R}) \cup \bigcup_{i=1}^n u_\lor^2(\qCQ_2^{i} \setminus \qCQ_2^{i'})$ with two points:
        \begin{itemize}
            \item $\homo'(u_\lor^1(\body{R})) = u_\lor^2(\body{R})$. Indeed, for any variable $x$ of $\body{R}$:
            \begin{itemize}
                \item  either $x \in \vars{\body{R}} \setminus \vars{\bigcup_{i=1}^n H_i''}$, so $\homo'(u_\lor^1(x)) = \homo'(x) = u_\lor^2(x)$ (because $u_\lor^1$ is a substitution from $\vars{\bigcup_{i=1}^n (\qCQ_1^{i'} \cup H_i'')}$);
                \item  or $x \in \fr{R} \cap \vars{\bigcup_{i=1}^n H_i''})$, so $\homo'(u_\lor^1(x)) = s(u_\lor^1(x)) = u_\lor^2(\homo(x)) = u_\lor^2(x)$ (because $\homo$ is a substitution from $\vars{\bigcup_{i=1}^n \qCQ_1^{i}}$\mic{ and recall that for any term $t$ in $P_{u_\lor^1}$, $s(u_\lor^1(t)) = u_\lor^2(\homo(t))$}).
            \end{itemize}
            

            \item \mic{$\homo'(u_\lor^1(\qCQ_i^{1} \setminus \qCQ_i^{1'})) \subseteq u_\lor^2(\qCQ_2^{i} \setminus \qCQ_2^{i'})$ for each $1\leq i\leq n$. In fact, we'll} show that $\homo'(u_\lor^1(\qCQ_1^{i} \setminus \qCQ_1^{i'})) = u_\lor^2(\homo(\qCQ_1^{i} \setminus \qCQ_1^{i'}))$ and since $\homo(\qCQ_1^{i} \setminus \qCQ_1^{i'}) \subseteq  \qCQ_2^{i} \setminus \qCQ_2^{i'}$ we'll be able to conclude. 
            To show that \mic{$\homo'(u_\lor^1(\qCQ_1^{i} \setminus \qCQ_1^{i'})) = u_\lor^2(\homo(\qCQ_1^{i} \setminus \qCQ_1^{i'}))$}, just see that for any \mic{$x\in \vars{\qCQ_1^{i} \setminus \qCQ_1^{i'}}$}:
            \begin{itemize}
                \item either 
                \mic{$x \in (\vars{\qCQ_1^{i'}} \cap \vars{\qCQ_1^i})$}, then $\homo'(u_\lor^1(x)) = s(u_\lor^1(x)) = u_\lor^2(\homo(x))$ ;
                \item or \mic{$x \in (\vars{\qCQ_1^{i}} \setminus \vars{\qCQ_1^{i'}})$}, then $\homo'(u_\lor^1(x)) = \homo'(x) = \homo(x) = u_\lor^2(\homo(x))$ (because $u_\lor^1$ is a substitution from $\vars{\bigcup_{i=1}^n (\qCQ_1^{i'} \cup H_i'')}$ and $u_\lor^2$ is a substitution from variables of $\bigcup_{i=1}^n (\qCQ_2^{i'} \cup H_i')$ and $\homo(x) \notin \vars{\bigcup_{i=1}^n (\qCQ_2^{i'} \cup H_i')}$).
            \end{itemize}
        \end{itemize}

\end{itemize}
\end{proof}


The following propositions \ref{prop-beta-reverse-alpha} and \ref{prop-alpha-reverse-beta} correspond to \textbf{Lemma \ref{lemma-compo}} (Point 1 and Point 2, respectively) in the paper.

\begin{proposition}\label{prop-beta-reverse-alpha}

Let a fact base $\fb$, a disjunctive rule $R$, a trigger $(R,\homo)$ on $\fb$ and let $\qUCQ$ be the UCQ $\alpha_\lor(\fb, R, \homo)$. Then there exists a disjunctive piece-unifier $\mu_\lor$ of $\qUCQ$ with $R$ such that $\fb \models \beta_\lor(\qUCQ, R, \mu_\lor)$.
\end{proposition}

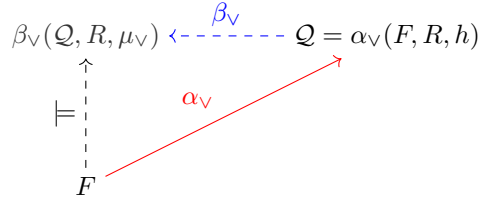
\begin{figure}[ht]
\begin{center}
\begin{tikzpicture}
\node (fb) at (0,0) {$\fb$};
\node (alphafb) at (4, 2) {$\qUCQ = \alpha_\lor(\fb, R, \homo)$};
\node (betaQ) at (0,2) {\color{darkgray}$\beta_\lor(\qUCQ, R, \mu_\lor)$};
\graph  {
  (fb) ->[red, "$\alpha_\lor$"] (alphafb),
  (alphafb) ->[dashed, blue, sloped, "$\beta_\lor$"] (betaQ),
  (fb) ->[dashed, "$\models$"]  (betaQ)
  
};
\end{tikzpicture}
\end{center}
\caption{\mic{Corresponding application of $\beta_\lor$ to the UCQ obtained by $\alpha_\lor$  is entailed by the original factbase (Prop. \ref{prop-beta-reverse-alpha})}}
\end{figure}

\begin{proof}

\mic{Let $\qUCQ = \alpha_\lor(\fb, R, \homo) = \{ \qCQ_i =  \fb \, \cup\, \homo^{{safe}_i}(\headi{i}{R}) ~|~ 1 \leq i \leq n\}$.
We build} $\mu_\lor = \{\mu_1, \dots, \mu_n\}$ a disjunctive piece-unifier as follows: for $1 \leq i \leq n$, $\mu_i = (\qCQ_i', \headi{i}{R}, P_{u_i})$ with $\qCQ_i' = \rho_i\circ \homo^{safe_i}(\headi{i}{R})$  ($\rho_i$ being a safe renaming of $\qCQ_i$) and $P_{u_i}$ the partition induced by $\rho_i \circ \homo^{safe_i}$ on $\terms{\qCQ_i'} \cup \terms{\headi{i}{R}}$.

First, we show that each $\mu_i$ is a piece-unifier of $\rho_i(\qCQ_i)$ with $\body{R}\rightarrow \headi{i}{R}$:
\begin{itemize}
    \item $\qCQ_i' \subseteq \rho_i(\qCQ_i)$ \mic{because $\homo^{{safe}_i}(\headi{i}{R})\subseteq \qCQ_i$ and $\qCQ_i'= \rho_i(\homo^{{safe}_i}(\headi{i}{R})$};
    \item $P_{u_i}$ the partition induced by $\rho_i \circ \homo^{safe_i}$ is admissible (thanks to Proposition \ref{prop-sub-adm}); 
    \item \mic{any $u_i$ associated with $P_{u_i}$ is such that} $u_i(\headi{i}{R}) = u_i(\qCQ_i')$ (thanks to Proposition \ref{prop-sub-unif});
    \item for each existential variable $z$ from $\headi{i}{R}$ we have $P_{u_i}[z] = \{z, \rho_i \circ \homo^{safe_i}(z)\}$ and $\rho_i \circ \homo^{safe_i}(z)$ is not a separating variable because $z$ is safely renamed twice, first by $.^{safe_i}$ and secondly by $\rho_i$.
\end{itemize}

Then, we show that the partition $P_{u_\lor} = join(\{P_{u_1}, \dots, P_{u_n}\})$ is admissible.

Since each $P_{u_i}$ is admissible, the non-admissibility of their join would be only due to a variable that appears in two classes with different constants from two partitions. The only variables that can be shared between two partitions \mic{of a set of piece-unifiers build from safe copies of CQs are the frontier variables of the considered disjunctive rule}.
But if a frontier variable shared by two $\headi{i}{R}$ is mapped on a constant, then it is mapped on the same constant \mic{because each $P_{u_i}$ is induced by $\rho_i \circ \homo^{safe_i}$ and only $\homo$ can send a variable to a constant.}

\mic{$\mu_\lor$ is therefore a disjunctive piece-unifier from $\qUCQ$ with $R$. Let $u_\lor$ be a substitution associated with $P_{u_\lor}$.}

Let $\fb' = 
\beta_\lor(\qUCQ, R, \mu_\lor) = u_\lor(B) \cup \bigcup\limits_{1 \leq i \leq n} u_\lor(\rho_i(\qCQ_i)\setminus \qCQ_i') =$ $u_\lor(B) \cup \bigcup\limits_{1 \leq i \leq n} u_\lor(\rho_i(\fb\mic{\,\cup\,\homo^{{safe}_i}(\headi{i}{R})})\setminus (\rho_i \circ \homo^{safe_i})(\headi{i}{R}))$ $\subseteq$ $u_\lor(B) \cup \bigcup\limits_{1 \leq i \leq n} u_\lor(\rho_i(\fb))$ \mic{(note this inclusion is not a simple equality because $\fb\,\cap\,\homo^{{safe}_i}(\headi{i}{R})$ can be not-empty)}.

We have just to observe  that $\rho_1^{-1} + \dots + \rho_n^{-1}+\homo$ is a homomorphism from  $u_\lor(B) \cup \bigcup\limits_{1 \leq i \leq n} u_\lor(\rho_i(\fb))$ to $\fb$:
\begin{itemize}
    \item $(\rho_1^{-1} + \dots + \rho_n^{-1}+\homo)(u_\lor(\rho_i(\fb))) = \fb$, indeed:
        \begin{itemize}
            \item If $\fb$ contains only constants, it is straightforward;
            \item If $\fb$ contains some variables, then they were renamed in $\rho_i(\fb)$. $u_\lor$ can only maps variables into two distinct sets of terms:
                \begin{itemize}
                    \item Assume that a variable of $\rho_i(\fb)$ is mapped to a variable in $\terms{\qCQ'_i}$. Then, $\rho_i^{-1}$ allows to recover the initial variable that was in $\fb$ (because no variable of $\fb$ can be in the same class of $P_{u_i}$ as an existential variable of $R$ and the other variables come from the application of $\rho_i$);
                    \item Otherwise, assume it is mapped to a variable in $\terms{\headi{i}{R}}$. Then, $h$ allows to recover the initial variable in $\fb$ (since these variables can only appear in $\qCQ_i$ through the frontier variables of $R$ thanks to the application of $h$ on $\headi{i}{R}$). 
                \end{itemize}
        \end{itemize}
    \item $(\rho_1^{-1} + \dots + \rho_n^{-1}+\homo)(u_\lor(B)) = \homo(B) \subseteq \fb$, indeed, by a similar reasoning:
        \begin{itemize}
            \item Assume that a variable in $B$ is sent by $u_\lor$ to a variable in $\terms{\qCQ'_i}$, then $\rho_i^{-1}$ allows to recover the variable in $\fb$ to which $h$ maps this variable from $B$;
            \item Assume it is mapped by $u_\lor$ to a variable in $\terms{\headi{i}{R}}$, then we simply have a variable in the domain of $h$ since it can only be a frontier variable.
        \end{itemize}
\end{itemize}

Since $\fb' \subseteq u_\lor(B) \cup \bigcup\limits_{1 \leq i \leq n} u_\lor(\rho_i(\fb))$, it follows that $\rho_1^{-1} + \dots + \rho_n^{-1}+\homo$ maps $F'$ to $\fb$. 

\end{proof}

\begin{proposition}\label{prop-alpha-reverse-beta}
Let $\qUCQ$ be a UCQ, $R$ be a disjunctive rule, $\mu_\lor$ be a disjunctive piece-unifier of $\qUCQ$ with $R$ and $\fb$ be the fact base $\beta_\lor(\qUCQ, R, \mu_\lor)$. Then, there exists a trigger $(R,\homo)$ on $\fb$ such that $\alpha_\lor(\fb, R, \homo) \models \qUCQ$.
\end{proposition}

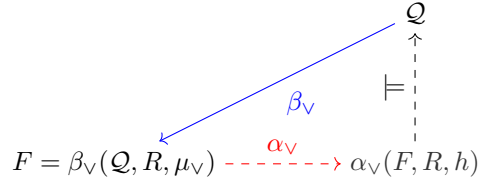
\begin{figure}[ht]
\begin{center}
\begin{tikzpicture}
\node (fb) at (0,0) {$\fb = \beta_\lor(\qUCQ, R, \mu_\lor)$};
\node (alphafb) at (4, 0) {\color{darkgray}$\alpha_\lor(\fb, R, \homo)$};
\node (Q) at (4,2) {$\qUCQ$};
\graph  {
  (fb) ->[dashed, red, "$\alpha_\lor$"] (alphafb) ->[dashed, "$\models$"] (Q),
  (Q) ->[blue, "$\beta_\lor$"] (fb)
};
\end{tikzpicture}
\end{center}
\caption{\mic{Corresponding application of $\alpha_\lor$ to the CQ obtained by $\beta_\lor$  entails the original UCQ (Prop. \ref{prop-alpha-reverse-beta})}}
\end{figure}

\begin{proof}

Let $\mu_\lor = \{\mu_1,\dots,\mu_n\}$, 
\mic{and let $u_{\lor}$ be a substitution associated with $join(\{P_{u_1}, \ldots, P_{u_n}\})$ with each $P_{u_i}$ being the partition in each $\mu_i$.
$(R,u_{\lor})$ is a trigger on $\fb= \beta_\lor(\qUCQ, R, \mu_\lor)$ since $u_{\lor}(\body{R}) \subseteq \fb$.} Let $\mathcal{F'} = \alpha_\lor(\fb, R, u_\lor) = \{ \fb'_i =  \fb \, \cup\, u_\lor^{{safe}_i}(\headi{i}{R}) ~|~ 1 \leq i \leq n\}$.

To prove that $\mathcal{F'} \models \qUCQ$, we'll show that for each $\fb_i'$, 
\mic{the CQ $\qCQ_i$ that is a safe copy of a CQ in $\qUCQ$ and was unified by $\mu_i$ with $\headi{i}{R}$} maps to $\fb_i'$ by the homomorphism $u_{\lor}^{safe_i}$.
\mic{ Then, let $\rho_i$ be the renaming substitution that produced $\qCQ_i$ from a CQ $Q\in \qUCQ$, we'll have $u_{\lor}^{safe_i}\circ \rho_i$ is a homomorphism from this $Q$ to $\fb_i'$. Thus by Proposition \ref{prop-specialization-set-fact-bases-homomorphism}, we can conclude that $\mathcal{F'} \models \qUCQ$.}

\mic{Let's now show that $u_{\lor}^{safe_i}$ maps $\qCQ_i$ to $\fb'_i$:} 
\begin{itemize}
    \item $u_{\lor}$ maps $\qCQ_i\setminus \qCQ'_i$ into $u_{\lor}(\qCQ_i\setminus \qCQ'_i) \mic{\,\subseteq \fb \subseteq \fb'_i}$ \mic{ and since $u_{\lor}^{safe_i}$ is an extension of $u_{\lor}$ to the existential variables of $R$, $u_{\lor}^{safe_i}(\qCQ_i\setminus \qCQ'_i)=u_{\lor}(\qCQ_i\setminus \qCQ'_i)$, so $u_{\lor}^{safe_i}$ maps $\qCQ_i\setminus \qCQ'_i$ into $\fb'_i$.}
    \item $u_{\lor}^{safe_i}$ maps $\qCQ'_i$ into $u_{\lor}^{safe_i}(\headi{i}{R}) \mic{\,\subseteq \fb'_i}$ because first \mic{$u_{\lor}$ unifies $\qCQ'_i$ and $H_i'\subseteq\headi{i}{R}$, \textit{i.e.} $u_{\lor}(\qCQ_i')=u_{\lor}(H'_i)$}, and second $.^{safe_i}$ maps $u_\lor(\headi{i}{R})$ into $u_\lor^{safe_i}(\headi{i}{R})$;
\end{itemize}

\end{proof}

\begin{lemma}
[Backward-forward Lemma]\label{lemma-back-forward}
Let $\fb$ be a fact base, $\qUCQ$ be a UCQ and $R$ be a disjunctive rule. For any disjunctive piece-unifier $\mu_\lor$ of $\qUCQ$ with $R$, if $\fb \models \beta_\lor(\qUCQ, R, \mu_\lor)$ then there is a trigger $(R,\homo)$ on $\fb$ such that $\alpha_\lor(\fb, R, \homo) \models \qUCQ$.
\end{lemma}

\begin{proof}
Thanks to the Proposition \ref{prop-alpha-reverse-beta}, we know that \mic{there is a trigger $(R,\homo)$ on $\fb_2 = \beta_\lor(\qUCQ, R, \mu_\lor)$ such} that $\alpha_\lor(\fb_2, R, \homo) \models \qUCQ$. Then, from Proposition \ref{prop-alpha-specialization}, we know that if $\fb \models \fb_2$, then we have $\alpha_\lor(F, R, \homo) \models \alpha_\lor(\fb_2, R, \homo)$. And thus, $\alpha_\lor(F, R, \homo) \models \qUCQ$ which is what we wanted to prove.
\end{proof}

\begin{lemma}
[Forward-backward Lemma]\label{lemma-forward-back}
Given any trigger $(R,\homo)$ on $\fb$,  if $\alpha_\lor(\fb, R, \homo) \models \qUCQ$ then either $\fb \models \qUCQ$ or there is a disjunctive piece-unifier $\mu_\lor$ of $\qUCQ$ with $R$, such that $\fb \models \beta_\lor(\qUCQ, R, \mu_\lor)$.
\end{lemma}

\begin{proof}
   Thanks to Proposition \ref{prop-beta-reverse-alpha}, we know that \mic{there is a disjunctive piece-unifier $\mu_\lor^2$ of $\qUCQ_2 = \alpha_\lor(\fb, R, \homo)$ with $R$ such that} $\fb \models \beta_\lor(\qUCQ_2, R, \mu_\lor^2)$. Then, from Proposition \ref{prop-beta-specialization}, we know that if $\qUCQ_2 \models \qUCQ$, either $\beta_\lor(\qUCQ_2, R, \mu_\lor^2) \models \qUCQ$ or there exists $\mu_\lor$ such that $\beta_\lor(\qUCQ_2, R, \mu_\lor^2) \models \beta_\lor(\qUCQ, R, \mu_\lor)$. Since $\fb \models \beta_\lor(\qUCQ_2, R, \mu_\lor^2)$, we have either $\fb \models \qUCQ$ or $\fb \models \beta_\lor(\qUCQ, R, \mu_\lor)$, which was what we wanted to prove.
\end{proof}


\begin{corollary}[of Lemma \ref{lemma-forward-back}]\label{cor-forward-back}
Let $\mathcal{F}$ be a set of fact bases, $\fb \in \mathcal{F}$, $R$ a disjunctive rule and $(R,\homo)$ a trigger on $\fb$. Let $\mathcal{F}_1$ be the set of fact bases obtained by the immediate derivation of $(\mathcal{F},R)$ by the trigger $(R,\homo)$, \textit{i.e.} $\mathcal{F}_1 = \mathcal F \setminus \{\fb\} \cup \alpha_\lor(\fb,R,\homo)$. Then, if $\mathcal{F}_1 \models \qUCQ$, either $\mathcal{F} \models \qUCQ$ or there exists a unifier $\mu_\lor$ of $\qUCQ$ with $R$ such that $\mathcal{F} \models \mic{\{\beta_\lor(\qUCQ,R,\mu_\lor)\}} \cup \qUCQ$.
\end{corollary}

\begin{proof}
Since $\mathcal{F}\setminus\{\fb\} \models \qUCQ$, we just have to prove that either $F \models \qUCQ$ or $\fb \models \beta_\lor(\qUCQ,R,\mu_\lor)$, which is exactly Lemma \ref{lemma-forward-back}.
\end{proof}

\mic{
We extend the notion of disjunctive chase result to any derivation tree or derivation sequence. So we call \emph{derivation tree result} the set of fact bases $res(\mathcal{T}) = \{ \bigcup\limits_{v \in nodes(\gamma)} \lambda(v) ~|~ \gamma \in \Gamma(\mathcal{T})\}$ where $\mathcal{T}$ is any derivation tree and $\lambda$ its labeling function. Also, we call \emph{derivation sequence result} the set of fact bases $res(\mathcal{D}) = \mathcal{F}_n$ where $\mathcal{F}_n$ is the last set of fact bases in the derivation $\mathcal{D}$.}

Note that if $\mathcal{T}$ is finite, we have $res(\mathcal{T}) \equiv res(\mathcal{D})$  for any derivation $\mathcal{D}$ that we can assign to $\mathcal{T}$. Indeed, for each finite sequence $\mathcal D_n$ of length $n$, $\mathcal F_n$ corresponds exactly to the labels of the leaves of a derivation tree built from the same trigger applications: hence, $\mathcal F_n$ is isomorphic to $res(\mathcal T)$. 

\begin{lemma}\label{lemma-finite-fact-base}
\mic{Let $\qCQ$ be a CQ, $\mathcal{T}=(V,E,\lambda)$ be a derivation tree, $\gamma$ be a branch of $\mathcal{T}$ and $\fb_\gamma \in res(\mathcal{T})$ the set of facts associated with $\gamma$, \textit{i.e.} $\fb_\gamma=\bigcup\limits_{v \in nodes(\gamma)} \lambda(v)$. If a homomorphism $\homo$ maps $\qCQ$ to $\fb_\gamma$
, then there is a vertex $v\in \gamma$ such that $\lambda(v)\models \qCQ$.}
\end{lemma}

\begin{proof}
To each atom of $\fb_\gamma$, we give a rank that corresponds to the depth\footnote{\mic{The depth of a vertex $v$ is defined as the length of the path from the root to $v$.}} of the \mic{vertex of $\gamma$} where it was produced.
Since $\homo(\qCQ)$ is finite, let $k$ be the maximum rank of the atoms in $\homo(\qCQ)$. \mic{Let $v$ be the vertex} at depth $k$ in \mic{$\gamma$, we have $\homo(\qCQ)\subseteq\lambda(v)$, so $\lambda(v)\models \qCQ$}.
\end{proof}

\begin{theorem} \label{th-finite-derivation-tree}
Let a UCQ $\qUCQ$, a set of disjunctive rules $\mathcal{R}$ and a fact base $\fb$. Then $\chase{\fb}{\mathcal{R}} \models \qUCQ$ iff there exists a finite derivation tree $\mathcal{T}$ of $(F,\mathcal{R})$ such that $res(\mathcal{T}) \models \qUCQ$.
\end{theorem}

%

\begin{proof}
$(\Leftarrow)$ We only need to 
\mic{extend the derivation tree ${\mathcal T}$}, in a fair way, to add what is missing in the tree. Indeed, by definition of the result of a \mic{derivation tree / disjunctive chase, each fact base of $\chase{\fb}{\mathcal{R}}$ includes at least one fact base of $res(\mathcal{T})$, so $\chase{\fb}{\mathcal{R}}\models res(\mathcal{T})$ and thus $\chase{\fb}{\mathcal{R}}\models \qUCQ$.}

$(\Rightarrow)$ Let $\mathcal{T}_C = (V, E, \lambda)$ the \mic{fair} derivation tree used to define $\chase{\fb}{\mathcal{R}}$, \mic{\textit{i.e.} $\chase{\fb}{\mathcal{R}} = \{ \fb_\gamma = \bigcup\limits_{v \in \textit{nodes}(\gamma)} \lambda(v) ~|~ \gamma \in \Gamma(\mathcal{T}_C)\}$.
For each $F_\gamma\in \chase{\fb}{\mathcal{R}}$, let $\qUCQ_\gamma\subseteq\qUCQ$ the set of CQs that maps to $F_\gamma$ ($\qUCQ_\gamma$ contains at least one CQ, cf. Proposition \ref{prop-specialization-set-fact-bases-homomorphism}). By lemma \ref{lemma-finite-fact-base}, for each CQ in $\qUCQ_\gamma$ there is a vertex $v\in \gamma$ such that $\lambda(v) \models \qCQ_\gamma$.
In each branch $\gamma$, we select $v_\gamma$ the highest of these vertices in $\gamma$.}

These selected vertices are called the terminal vertices. We build the subtree $\mathcal{T}'$ of $\mathcal{T}_C$ by deleting from $\mathcal{T}_C$ all the vertices that are successors of a terminal vertex. Thus every branch of $\mathcal{T}'$ is finite. We show that (1) $\mathcal{T}'$ is still a derivation tree and (2) it is finite. 

\begin{enumerate}
    \item By construction, each node in $\mathcal{T}'$ is either a terminal node (in which case, it is a leaf), or we did not erase any of its children (and so, its children still correspond to the result of applying a trigger). Thus, $\mathcal{T}'$ is still a derivation tree.
    \item Since each rule is finite, each node in a derivation tree has a finite number of children (it is locally finite). According to König's infinity Lemma \cite{KONIG27}, ``an infinite, locally finite rooted tree has an infinite branch''. Its contrapositive is ``a locally finite rooted tree with no infinite branch is finite''. Thus, $\mathcal{T}'$ is finite.
\end{enumerate}
\end{proof}

\begin{corollary}[of Theorem \ref{th-finite-derivation-tree}]\label{cor-finite-derivation-tree}
$F, \mathcal{R} \models \qUCQ$ iff there exists a finite derivation tree $\mathcal{T}$ of $(F,\mathcal{R})$ such that $res(\mathcal{T}) \models \qUCQ$. 
Equivalently, $F, \mathcal{R} \models \qUCQ$ iff there exists a derivation $\mathcal{D}$ from $F$ with $\mathcal{R}$ such that $res(\mathcal{D}) \models \qUCQ$.
\end{corollary} 


\paragraph{Theorem \ref{th-sound-complete-disjunctive-rewriting}}
Let $(\fb, \mathcal R)$ be a disjunctive KB and $\qUCQ$ be a (Boolean) UCQ. Then, $\fb, \mathcal{R} \models \qUCQ$ iff there is a piece-rewriting $\qUCQ'$  of $\qUCQ$ such that $\fb\models \qUCQ'$. 

\begin{proof} We show that there exists a derivation of $(\mic{\{\fb\}}, \mathcal R)$ leading to an $\mathcal F_i$ such that $\mathcal F_i \models \qUCQ$ iff there exists a piece-rewriting \mic{$\qUCQ'$} of $\qUCQ$ with $\mathcal R$ such that \mic{$\fb\models \qUCQ'$.}  

$(\Rightarrow)$ We prove the first direction by induction on the number of rule applications in a 
\mic{derivation sequence} $\mathcal{D}$ such that $\qUCQ$ maps to $res(\mathcal{D})$ (such a tree / derivation exists: see Corollary \ref{cor-finite-derivation-tree}).

At rank $0$, the property is trivially true by taking $\qUCQ'=\qUCQ$. 
Let us assume that it is true at rank $n$. Let $\mathcal{D} = (\mathcal{F}_0 = \{F\}) \xrightarrow{t_1}{} \dots\xrightarrow{t_n}{}\mathcal{F}_n \xrightarrow{(R,h)}{}\mathcal{F}_{n+1}$ with $\qUCQ$ that maps to $\mathcal{F}_{n+1}$. 
By using the Corollary \ref{cor-forward-back}, we have  either: 
\begin{enumerate}
    \item $\mathcal F_n \models \qUCQ$;
    \item or there exists $\mu_\lor$ such that $\mathcal F_n \models \mic{\{\beta_\lor(\qUCQ, R, \mu_\lor)\}} \cup \qUCQ$.
\end{enumerate}

In both cases, we have a UCQ that  maps to $\mathcal F_n$. Let us name it $\qUCQ_n$.
By induction hypothesis, there exists a piece-rewriting $\qUCQ'$ of $\qUCQ_n$ such that \mic{$\fb \models \qUCQ'$. By definition, $\qUCQ_n$ is a one-step piece-rewriting of $\qUCQ$, and thus $\qUCQ'$ is also a piece-rewriting of $\qUCQ$.}

\smallskip
$(\Leftarrow)$ We prove the opposite direction by induction on the length of the rewriting sequence \mic{producing $\qUCQ'$} from $\qUCQ$ and relying upon Lemma \ref{lemma-back-forward}. The property is trivially true at rank $0$ by taking $\mathcal{F}_0=\{\fb\}$. Let us assume it is true at rank $n$. Assume that $\qUCQ_{n+1}$ is obtained from $\qUCQ$ by a rewriting sequence $\qUCQ = \qUCQ_0, \qUCQ_1, \dots, \qUCQ_n, \qUCQ_{n+1} = \beta_\lor(\qUCQ_n, R, \mu_{\lor}) \cup \qUCQ_n$ of length $n+1$, and $\fb\models\qUCQ_{n+1}$. \mic{So there is a CQ $\qCQ$ in $\qUCQ_{n+1}$ such that $\fb\models\qCQ$.} 
We have two cases:
\begin{enumerate}
    \item $\qCQ \in \qUCQ_n$: then, by induction hypothesis, there exists $\mathcal{F}_i$ such that $\mathcal{F}_i \models \qCQ$, \mic{thus $\mathcal{F}_i \models \qUCQ_n$ and also $\mathcal{F}_i \models \qUCQ_{n+1}$}.
    \item $\qCQ = \beta_\lor(\qUCQ_n, R, \mu_{\lor})$: then, by Lemma \ref{lemma-back-forward}, there exists $\mathcal{F}_1 = \alpha_\lor(\fb, R, \homo)$ such that $\mathcal{F}_1 \models \mathcal{Q}_n$. So, we have that for each $\fb_m \in \mathcal{F}_1$, $\fb_m \models \mathcal{Q}_n$. And by induction hypothesis, it holds that for each $\fb_m$, there exists \mic{a derivation of $(\{\fb_m\}, \mathcal R)$ leading to a }$\mathcal{F}_m$ such that $\mathcal{F}_m \models \qUCQ$ and thus we have a derivation from $\mathcal{F}_1$ that produces $\mathcal{F}_i$ (that is the union of all $\mathcal{F}_m$) such that $\mathcal{F}_i \models \qUCQ$.
\end{enumerate}
\end{proof}
\newpage
\section{Proofs of Section \ref{sec-fus}}

\textbf{Theorem \ref{th-non-fus}.}
\emph{ 
Let $R = B \rightarrow H_1 \lor H_2$ be a source-to-target rule that is not disconnected nor equivalent to a conjunctive rule. 
Then, there is a CQ $\qCQ$ such that  $(\{\qCQ\},\{R\})$ is not UCQ-rewritable. }

\begin{proof}
Let $R = B[\vect{x_1}, \vect{x_2}, \vect{y}] \rightarrow \exists  \vect{z_1}~H_1[ \vect{x_1},  \vect{z_1}] \lor \exists  \vect{z_2}~H_2[ \vect{x_2},  \vect{z_2}]$, where:
\begin{itemize}
\item $\fr{R} =  \vect{x_1} \cup  \vect{x_2}$; $ \vect{x_1}$ and $ \vect{x_2}$ may share variables;
\item $ \vect{x_i} \neq \emptyset$ ($i = 1,2$) since $R$ is not disconnected. 
\end{itemize}
We build the following Boolean CQ:


$$\qCQ = \{ H^s_1[ \vect{v_1},  \vect{w_1}], p( \vect{v_1},  \vect{v_2}), H^s_2[ \vect{v_2},  \vect{w_2}] \}$$

where each $H^s_i[ \vect{v_i},  \vect{w_i}]$ is a safe copy of $H_i[ \vect{x_i},  \vect{z_i}]$ and $p$ is a fresh predicate. 
Note that, since $R$ is connected, both $H_1$ and $H_2$ have a frontier variable, and frontier variables being safely renamed in each $H^s_i$, we have $\vect{v_1} \cap \vect{v_2} = \emptyset$, hence the arity of $p$ is at least $2$. 
In $p( \vect{v_1},  \vect{v_2})$ the order on the variables is important: a fixed order is chosen on $ \vect{x_i}$ (hence, $ \vect{v_i}$) and the tuple $ \vect{v_1}$ comes before the tuple $ \vect{v_2}$. Hence, $p( \vect{v_1}, \vect{v_2})$ can be seen as ``directed'' from  $\vect{v_1}$ to $\vect{v_2}$. 
We then proceed in two steps.
\begin{enumerate}
\item We show that we can produce an infinite set $\qUCQ$ whose element CQs are pairwise incomparable by homomorphism. 
Let $\qCQ_0 = \qCQ$. 
At each step $i \geq 1$, $\qCQ_i$ is produced from a safe copy of $\qCQ$ unified with $H_1$ and a safe copy of $\qCQ_{i-1}$ unified with $H_2$. The piece-unifiers unify $H_1^s$ (resp. $H_2^s$) in $\qCQ$ (resp. $\qCQ_{i-1}$) according to the isomorphism from $H_1^s$ (resp. $H_2^s$) to $H_1$ (resp. $H_2$). 
Any CQ $\qCQ_k$ in $\qUCQ$ is connected and follows the ``pattern'' $H^s_1.p.(B.p)^k.H^s_2$, where occurrences of $p$-atoms all have the same direction; hence, two ``adjacent'' $p$-atoms, i.e., that share variables with the same copy $B_i$ of a $B$, cannot be mapped one onto the other (by a homomorphism that maps $B_i$ to itself).  
\item We show that no CQ $\qCQ'$ that can be produced by piece-rewriting maps by homomorphism to a CQ from $\qUCQ$, except by isomorphism. 
When there is no (conjunctive) piece-unifier that unifies  $H_1[ \vect{v_1},  \vect{w_1}]$ in $\qCQ$ with $H_2[ \vect{x_2},  \vect{z_2}]$ (then, the same holds if we exchange $H_1$ and $H_2$), all the produced $\qCQ'$ are more specific than (including isomorphic to) CQs from $\qUCQ$. 
Otherwise, assume that a CQ $Q'$ is produced by unifying $H_1[ \vect{v_1},  \vect{w_1}]$ with $H_2[ \vect{x_2},  \vect{z_2}]$. If $Q'$ can be mapped by homomorphism to a $Q_n \in \mathcal Q$, the arguments of any $p$-atom in $Q'$ must be pairwise distinct variables. We show that it leads to have $R$ equivalent to the conjunctive rule $B \rightarrow H_i$ (with $i =1$ or $i=2$), which contradicts the hypothesis on $R$. 
%
\end{enumerate}
It follows that $\qUCQ$ is a subset of any sound and complete rewriting of $\{\qCQ\}$ with $\{R\}$, hence the pair $(\{\qCQ\},\{R\})$ does not admit a UCQ-rewriting.

\paragraph{Details on step 1.} We consider the infinite sequence $\qUCQ_0, \ldots, \qUCQ_i, \ldots$, where $\qUCQ_0 = \{ \qCQ_0 = \qCQ \}$ and for all $i > 0$, 
$\qUCQ_i = \qUCQ_{i-1} \cup \{ \qCQ_i \}$, where $\qCQ_i$ is obtained by a (disjunctive) piece-unifier that unifies safe copies of $\qCQ_0$ and $\qCQ_{i-1}$, with $H_1$ and $H_2$ respectively, according to the isomorphism from $H_1^s$ (resp. $H_2^s$) to $H_1$ (resp. $H_2$). By an easy induction on the length $k$ of the rewriting sequence leading to $\qUCQ_k$ ($k \geq 0$), we check that all the CQs $\qCQ_k$  are of the following form:


$$H^s_1[\vect{v_1^0},\vect{w_1}] \land p(\vect{v_1^0}, \vect{v_2^0}) \land 
\left(\bigwedge\limits_{i = 1}^k B[\vect{v_2^{i-1}},\vect{v_1^i}, \vect{y_i}] \land p(\vect{v_1^i}, \vect{v_2^i}) \right) 
\land H^s_2[\vect{v_2^k},\vect{w_2}]$$

Moreover,  $\qCQ_k$ is connected. Indeed, by hypothesis, $R$ is connected, hence $B$ is connected, or we have $\vect{v_1^{i-1}} \cap \vect{v_2^i} \neq \emptyset$, for all $ i >0$, i.e., two $p$-atoms adjacent to a $B$ share a variable. 

Since the two $p$-atoms connected to an occurrence of $B$ are ``in the same direction'', they do not fold one onto the other. Hence, if a CQ $\qCQ_i$ maps to a CQ $\qCQ_j$ ($i\neq j$), it is necessarily by an injective homomorphism. However, this is impossible, because the ``chains'' that underlie these CQs are of different length while the 
copies of $H_1$ and $H_2$ at their extremities should be mapped one onto the other. Hence, the set $\qUCQ$ defined as the union of all the $\qUCQ_i$ for $i \in  \mathbb{N}$, is composed of pairwise incomparable CQs. 

\medskip

%
%

\paragraph{Details on step 2.} 

\textbf{(1)} We first consider the case where there is no (conjunctive) piece-unifier that unifies $H_1[ \vect{v_1},  \vect{w_1}]$ in $\qCQ$ with $H_2[ \vect{x_2},  \vect{z_2}]$ (then, the same holds if we exchange $H_1$ and $H_2$) and show that the produced CQs are more specific than (including isomorphic to) CQs from $\qUCQ$.  Indeed, in this case, all the CQs produced are of the above general form, except that the $p$-atoms may be specialized, as well as the $B$'s on their frontier (it is the case if we consider more specific unifiers than the ones used to build $\mathcal Q$).  Let us prove it by induction on the length $l$ of a rewriting sequence. This is true for $l=0$. Assume this is true until $l=n$. For $l=n+1$, let $\qUCQ_j$ and $\qUCQ_k$, with (in simplified form) $\qUCQ_j = H^s_1.p.(B.p)^j.H^s_2$ unified with $H_1$ and $\qUCQ_k = H^s_1.p.(B.p)^k.H^s_2$ unified with $H_2$. The produced CQ has the form $H^s_1.p.(B.p)^{j+k+1}.H^s_2$, hence it is more specific than $\qCQ_{j+k+1}$, as defined in the step 1 of the proof. 
\textbf{(2)} Otherwise, let $Q_k$ be a CQ produced by unifying $H^s_1[ \vect{v_1},  \vect{w_1}]$ with $H_2[ \vect{x_2},  \vect{z_2}]$ (if we exchange $H_1$ and $H_2$, the case is similar). 
If $Q_k$ can be mapped by homomorphism to a $Q_n \in \mathcal Q$, any $p$-atom in $Q_k$ must have pairwise distinct variables. Hence, when an atom set of the form $H_1$ is unified with an atom set of the form $H_2$, the (copies of the) frontier variables in each set have to remain distinct (i.e., no frontier variable can be unified with another frontier variable in the same set).
From this observation and the fact that two existential variables of $H_2$ cannot be unified together, there is a homomorphism 
from $H^s_1[ \vect{v_1},  \vect{w_1}]$ to $H_2[ \vect{x_2},  \vect{z_2}]$, with $\vect{v_1}$ mapped to $\vect{x_2}$. 
Since by construction of the rewriting, an $H^s_1$ is never specialized (by merging two variables or replacing a variable by a constant), $H^s_1$ is isomorphic to $H_1$ (with frontier variables mapped to frontier variables). Hence, there is a homomorphism $h$ from $H_1$ to $H_2$, with frontier variables mapped to frontier variables. Now, two cases: either $h(B)$ maps to $B$ by a homomorphism invariant on the frontier variables of $h(B)$, and 
 $B \rightarrow H_2 \models B \rightarrow H_1$,
 hence $R$ is equivalent to the conjunctive rule $B \rightarrow H_2$, which is excluded by hypothesis; or $h(B)$ does not map to $B$ by a homomorphism invariant on the frontier variables of $h(B)$, and it does not map to a $B$ by a homomorphism from  $Q_k$ to $Q_n$, hence there is no homomorphism from $Q_k$ to $Q_n$. 

\end{proof} 

%

\newpage
\section{Proofs of Section \ref{sec-disjunctive-mappings}}

Recall that \emph{disjunctive mapping rewritability} is the following problem: Given a set of disjunctive $\mathcal S$-to-$\mathcal T$-rules $\mathcal M$ and a UCQ $\qUCQ$ on $\mathcal T$, does the pair $(\qUCQ, \mathcal M)$ admit a UCQ-$\mathcal S$-rewriting?

\paragraph{Theorem \ref{th-map-rewrite-undec}} Disjunctive mapping rewritability is undecidable.

\medskip

To prove it, we build a reduction from the following problem: Given a Boolean CQ $\qCQ$ and a set of (conjunctive) datalog rules $\mathcal R$, does the pair $(\qCQ, \mathcal R)$ admit a UCQ-rewriting? This problem is undecidable, 
which follows from the undecidability of determining whether a datalog program is uniformly bounded \cite{GAIFMAN93}.
Indeed, a datalog program $\mathcal R$ is uniformely bounded if and only if the pair $(\qCQ, \mathcal R)$ is UCQ-rewritable for any full atomic query $\qCQ$, i.e, in which all the variables are answer variables. 
Since there is a finite number of non-isomorphic atomic CQs to consider, it follows that determining if a pair $(\qCQ, \mathcal R)$ is UCQ-rewritable for $\qCQ$ an atomic CQ is also undecidable. 
In turn, this problem can be reduced to the problem of determining whether a pair $(\qCQ', \mathcal R)$ is UCQ-rewritable for $\qCQ'$ a Boolean CQ. To build $\qCQ'$, we just add to $\qCQ$ an atom with special predicate \emph{answer} which contains all the variables of $\qCQ$. This ensures that answer variables are properly considered when comparing two generated CQs.

W.l.o.g. we assume that datalog rules have no constants (and an atomic head). 

\medskip

Our reduction translates each instance $(\qCQ, \mathcal R)$ of the conjunctive datalog UCQ-rewriting problem, defined on a set of predicates $\mathcal P$, into an instance $(\mathcal Q^{Q,\mathcal R},\mathcal M^{Q,\mathcal R})$ of the disjunctive mapping rewritability problem, defined on a pair of predicats sets $(\mathcal S, \mathcal T)$ such that:
\begin{itemize}
\item  $\mathcal S = \mathcal P\cup\{T\}$, where $T$ is a fresh unary predicate, 
\item  $\mathcal T$ is the union of: (1) a set of predicates in bijection with $\mathcal S$, where each predicate is topped with a hat (e.g. $\hat{p}$ is obtained from $p$), and (2) a set of fresh predicates in bijection with $\mathcal R$, where we denote by $p_{R_i}$ the predicate associated with the rule $R_i$; 
the arity of each $p_{R_i}$ is $|\fr{R_i}|$. 
\end{itemize}

We denote $A[\vect{x}, \vect{y}]$ a set of atoms that uses the variables in $\vect{x}$ and $\vect{y}$.
We also denote by $T[\vect{x}]$ the conjunction of atoms $T(x_i)$ for each $x_i \in \vect{x}$, i.e. $T[\vect{x}]=T(x_1) \land \dots \land T(x_n)$ where $|\vect{x}|=n$. Similarly, $\hat{T}[\vect{x}] = \hat{T}(x_1) \land \dots \land \hat{T}(x_n)$. 
Let $Q$ any CQ ( or set of atoms) on $\mathcal S$, we denote by $\hat Q$ the CQ (or set of atoms) $Q$ whose predicates have all been renamed with a hat, 
$\hat Q$ is thus on $\mathcal T$.
Let $Q$ any CQ, we denote by $Q^T$ the CQ $Q$ completed with a $T$ atom on each term. Then, $\hat Q^T$ is the CQ obtained from $\hat Q$ by adding its $T$ atoms. Finally,   $\widehat{Q^T}$ is obtained from $Q^T$ by substituting each predicate $p$ (including $T$) by $\hat{p}$. 

\medskip
\paragraph{Definition of the reduction}
Let a CQ $\qCQ = \exists \vect{x_\qCQ} ~ B_\qCQ[\vect{x_\qCQ}]$ and a datalog rule set $\mathcal R = \{R_1, \dots, R_n\}$ with each $R_i = B_i[\vect{x_i}, \vect{y_i}] \rightarrow H_i[\vect{x_i}]$. We define the UCQ $\qUCQ^{\qCQ,\mathcal R}$ and the disjunctive datalog mapping $\m^{\qCQ,\mathcal R}$ associated with $Q$ and $\mathcal R$ as follows: 
\begin{itemize}
\item $\qUCQ^{\qCQ,\mathcal R} = ~~  \{\qCQ_\qCQ\} \cup \qUCQ_{\mathcal R}$ with \\
    $~~~~~~~~\qCQ_Q = \exists \vect{x_\qCQ}. \hat {B_\qCQ}[\vect{x_\qCQ}] \land \hat T[\vect{x_\qCQ}]$, i.e.,  $\qCQ_Q = \widehat{\qCQ^T}$,\\
    $~~~~~~~~\mathcal {Q_R}=\{\qCQ_{R_i} = \exists \vect{x_i}, \vect{y_i}. \hat {B_i}[\vect{x_i}, \vect{y_i}] \land p_{R_i}(\vect{x_i}) \land \hat T[\vect{x_i}, \vect{y_i}] ~|~ R_i \in \mathcal R\}$
\item $\m^{\qCQ,\mathcal R} = \m_\mathcal R \cup \m_{trans}$ with \\
    $~~~~~~~~\m_\mathcal R =\{ m_{R_i} = T[\vect{x_i}] \rightarrow p_{R_i}(\vect{x_i}) \lor \hat H_i(\vect{x_i}) ~|~ R_i \in \mathcal R\}$ \\
    $~~~~~~~~\m_{trans} = \{p(\vect{x}) \rightarrow \hat p(\vect{x}) ~|~ p \in \mathcal S\}$
\end{itemize}

\medskip

Let us comment on the reduction. The UCQ $\qUCQ^{\qCQ,\mathcal R}$ is built from $Q$ and, for every $R_i \in \mathcal R$, a CQ $\qCQ_{R_i}$. Each $\qCQ_{R_i}$ is composed of the conjunction of $\body{R_i}$ and a special atom $p_{R_i}(\vect{x_i})$, where $p_{R_i}$ is a fresh predicate associated with $R_i$ and $\vect{x_i}$ is the frontier of $R_i$. The idea is that $p_{R_i}(\vect{x_i})$ will be unifiable (and thus erasable) only with a corresponding mapping assertion $m_{R_i}$,  which moreover enforces to have a CQ containing an atom unifiable with $\head{R_i}$. Then, for each term $t$ in a CQ, one adds a unary atom $T(t)$. The set of rules $\m^{\qCQ,\mathcal R}$ is built by creating, for each rule $R_i$, a disjunctive rule $m_{R_i}$ with a body that contains a $T(x)$ atom for each frontier variable $x$ of $R_i$, and a head with the special atom associated with $R_i$ as first disjunct, and $\head{R_i}$ as second disjunct. 
Finally, 
 the predicates from $\mathcal S$ of each atom in $\qUCQ^{\qCQ,\mathcal R}$ or in the head of disjunctive rules in $\m^{\qCQ,\mathcal R}$ are turned into target predicates (i.e., renamed with a ``hat"), and a set of atomic $\mathcal{S}$-to-$\mathcal{T}$ rules $\m_{trans}$ is added to translate the source predicates into target predicates.

\medskip

Note that there is a natural bijection (up to variable renaming) between the CQs defined on $\mathcal P$ and the CQs defined on $\mathcal S$ that have a $T$-atom on each term: to $Q_w$ on $\mathcal P$ we assign the CQ $Q_w^T$ composed of $Q_w$ (or any CQ isomorphic to $Q_w$) completed by $T$-atoms on each term.

\medskip
Then the correctness of the reduction is proved thanks to three lemmas:
\begin{itemize}
\item We prove in Lemma \ref{reduction-forward} that for any CQ $Q_w$ belonging to a piece-rewriting of $\{\qCQ\}$ with $\mathcal R$, a CQ isomorphic to $Q_w^T$ belongs to a piece-rewriting of $\mathcal Q^{\qCQ,\mathcal R}$ with $\m^{\qCQ, \mathcal R}$.
\item We prove in Lemma \ref{lemma-S-rewriting} that any CQ $Q_S$ belonging to an $\mathcal S$-rewriting of $\mathcal Q^{\qCQ,\mathcal R}$ with $\m^{\qCQ, \mathcal R}$ is of the form $Q_S = (Q_\mathcal P)^T$ where $Q_\mathcal P$ is a set of atoms on $\mathcal P$.
\item We prove in Lemma \ref{reduction-reverse} that for any CQ of the form $(Q_\mathcal P)^T$, with $Q_\mathcal P$ on $\mathcal P$, belonging to a piece-rewriting of $\mathcal Q^{\qCQ,\mathcal R}$ with $\m^{\qCQ, \mathcal R}$, a CQ isomorphic to $Q_\mathcal P$ belongs to a piece-rewriting of $\{\qCQ\}$ with $\mathcal R^\star$, where $\mathcal R^\star$ is the reflexive and transitive closure of $\mathcal R$ by unfolding. This lemma is established by showing that any CQ in a piece-rewriting of $\mathcal Q^{\qCQ,\mathcal R}$ with $\m^{\qCQ, \mathcal R}$ ``corresponds'' either to a rule from $\mathcal R^\star$, or to a piece-rewriting of $\{\qCQ\}$ with $\mathcal R^\star$. 
\end{itemize}

Furthermore, the proof implicitly uses the following observations:
\begin{enumerate}
\item Let $\mathcal Q$ be a finite rewriting of $\{Q\}$ with $\mathcal R$. Then there is a piece-rewriting $\mathcal Q_i$ of $\{Q\}$ with $\mathcal R$ such that  $\mathcal Q\models \mathcal Q_i$.\\
\textit{Proof}: For every complete rewriting $\mathcal Q'$ of $\{Q\}$ with $\mathcal R$, we have $\mathcal Q\models \mathcal Q'$ 
(indeed, let $M$ be a model of  $\mathcal Q$ and $h$ be a witnessing homomorphism from a CQ $Q'$ in $\mathcal Q$ to $M$. Let $F = h(Q')$. 
Since $F \models \mathcal Q$ and $\mathcal Q$ is sound, we have $\mathcal R, F \models Q$, hence  $F \models \mathcal Q'$ because $\mathcal Q'$ is complete. Hence, $M$ is a model of  $\mathcal Q'$). Since piece-rewriting is a complete procedure, there is a complete set of CQs $\mathcal Q_i$ produced by a possibly infinite sequence of piece-rewritings. Then, $\mathcal Q\models \mathcal Q_i$. This means that for each CQ $Q' \in \mathcal Q$, there is a CQ $Q_j \in \mathcal Q_i$ such that $Q' \models Q_j$. We can restrict $\mathcal Q_i$ to these $Q_j$ while keeping the entailment from  $\mathcal Q$. 
\item Let $\mathcal Q$ be a UCQ-rewriting of $\{Q\}$ with $\mathcal R$. Then there is a complete piece-rewriting $\mathcal Q_i$ of $\{Q\}$ with $\mathcal R$ such that  $\mathcal Q_i \equiv \mathcal Q$.\\
\textit{Proof}: Let $\mathcal Q_i$ be a complete set of CQs obtained by a possibly infinite sequence of piece-rewritings of $\{Q\}$ with $\mathcal R$. As previously, we consider a model $M$ of $\mathcal Q_i$ and $F$ a (finite)-witnessing subset of $M$. Since $\mathcal Q_i$ is sound, we have $\mathcal R, F \models Q$, hence $F \models \mathcal Q$ because $\mathcal Q$ is complete. Hence, $M$ is a model of $\mathcal Q$ and $\mathcal Q_i \models \mathcal Q$. We do the same reasoning by considering a model of $\mathcal Q$ to conclude that $\mathcal Q \models \mathcal Q_i$. We can restrict $\mathcal  Q_i$ to an equivalent finite subset because $\mathcal Q$ is finite and equivalent to $\mathcal  Q_i$ (see e.g., Theorem 1 in \cite{SWJ15}). 
\end{enumerate}

\begin{proof}[Proof of Theorem \ref{th-map-rewrite-undec}] 
We prove that there exists a UCQ-rewriting of $(\{Q\},\mathcal R)$ iff there exists a UCQ-$\mathcal S$-rewriting of $(\mathcal Q^{\qCQ,\mathcal R},\m^{\qCQ, \mathcal R})$.

($\Rightarrow$) Let $\mathcal Q$ be a UCQ-rewriting of $(\{Q\},\mathcal R)$. Then there exists a piece-rewriting $\mathcal Q_i$ of $\{Q\}$ with $\mathcal R$ such that $\mathcal Q_i\equiv \mathcal Q$.
By Lemma \ref{reduction-forward}, there is a piece-rewriting $\mathcal Q_j$ of $\mathcal Q^{\qCQ,\mathcal R}$ with $\m^{\qCQ, \mathcal R}$ that contains a subset of CQs in natural bijection with those in $\mathcal Q_i$. 
Let $\mathcal Q^{\mathcal S}_j$ be the subset of $\mathcal Q_j$ that contains only the CQs on $\mathcal S$. $\mathcal Q^{\mathcal S}_j$ is a finite $\mathcal S$-rewriting of $(\mathcal Q^{\qCQ,\mathcal R},\m^{\qCQ, \mathcal R})$.

Suppose $\mathcal Q^{\mathcal S}_j$  is not a complete $\mathcal S$-rewriting. Then, by Lemma \ref{lemma-S-rewriting}, there is a CQ $(Q_\mathcal P)^T$ which belongs to an $\mathcal S$-rewriting of $(\mathcal Q^{\qCQ,\mathcal R},\m^{\qCQ, \mathcal R})$ but which is not more specific than any of the CQs in $\mathcal Q^{\mathcal S}_j$. 
Then there is a CQ $(Q'_\mathcal P)^{T}$ that belongs to a piece-rewriting $\mathcal Q_j'$ of $\mathcal Q^{\qCQ,\mathcal R}$ with $\m^{\qCQ, \mathcal R}$ such that $(Q_\mathcal P)^{T} \models (Q'_\mathcal P)^{T}$.
Then by Lemma \ref{reduction-reverse}, $Q'_\mathcal P$ is isomorphic to a CQ belonging to a piece-rewriting of $\{\qCQ\}$ with $\mathcal R^\star$, hence	
to a rewriting of $(\{Q\},\mathcal R)$. Since 
$\mathcal Q_i$ is a UCQ-rewriting, there is a $Q_c$ in  $\mathcal Q_i$ such that $Q'_\mathcal P\models Q_c$. Hence, $(Q'_\mathcal P)^{T}\models (Q_c)^T$ (and thus $(Q_\mathcal P)^{T}\models (Q_c)^T$) and, since $(Q_c)^T$ belongs to $\mathcal Q^{\mathcal S}_j$ , this contradicts  the assumption that  
$(Q_\mathcal P)^T$ is not more specific than a CQ in $\mathcal Q^{\mathcal S}_j$.

\smallskip
($\Leftarrow$) Let $\mathcal Q^{\mathcal S}$ be a UCQ-$\mathcal S$-rewriting of $(\mathcal Q^{\qCQ,\mathcal R},\m^{\qCQ, \mathcal R})$.
Then there exists a piece-rewriting $\mathcal Q_i$ of $\mathcal Q^{\qCQ,\mathcal R}$ with $\m^{\qCQ, \mathcal R}$ such that $\mathcal Q^{\mathcal S} \models \mathcal Q_i$ (i.e., for each CQ $Q'$ in $\mathcal Q^{\mathcal S}$, there is a CQ $Q''$ in $ \mathcal Q_i$ such that $Q' \models Q''$).
Consider $\mathcal Q^{\mathcal S}_i$ the subset of $\mathcal Q_i$ that contains only the CQs on $\mathcal S$.  
We still have
 $\mathcal Q^{\mathcal S} \models \mathcal Q^{\mathcal S}_i$. Since $\mathcal Q^{\mathcal S}$ is complete w.r.t. $\mathcal S$, so is $\mathcal Q^{\mathcal S}_i$. Thus $\mathcal Q^{\mathcal S} \equiv \mathcal Q^{\mathcal S}_i$.  
%
%
%
%
By Lemma \ref{lemma-S-rewriting}, any CQ in $\mathcal Q^{\mathcal S}_i$ is of the form $(Q_\mathcal P)^T$ as required in Lemma \ref{reduction-reverse}. 
So, by Lemma \ref{reduction-reverse}, there is a piece-rewriting $\mathcal Q_j$ of $\{\qCQ\}$ with $\mathcal R^\star$ that contains all the CQs in natural bijection with those in $\mathcal Q^\mathcal S_i$. 
So $\mathcal Q_j$ is a finite rewriting of $\{\qCQ\}$ with $\mathcal R^\star$. 
Since $\mathcal R^\star \equiv \mathcal R$ (see also Proposition \ref{prop-unfolding}), it is also a finite rewriting of $\{\qCQ\}$ with $\mathcal R$.

Suppose $\mathcal Q_j$ is not complete. Then there is a CQ $Q_\mathcal P$ that belongs to a rewriting of $(\{\qCQ\},\mathcal R)$ and is not more specific than any of the  CQs in $\mathcal Q_j$.
Then there is a CQ $Q'_\mathcal P$ that belongs to a piece-rewriting 
of $\{\qCQ\}$ with $\mathcal R$ such that $Q_\mathcal P \models Q'_\mathcal P$.
Then by Lemma \ref{reduction-forward}, $(Q'_P)^{T}$ is isomorphic to a CQ belonging to a piece-rewriting of 
$\mathcal Q^{\qCQ,\mathcal R}$ with $\m^{\qCQ, \mathcal R}$.  
Since $\mathcal Q^\mathcal S_i$ is complete w.r.t. $\mathcal S$, there is a $(Q_c)^T$ in  $\mathcal Q^\mathcal S_i$ such that $(Q'_\mathcal P)^{T}\models (Q_c)^T$. We also have $Q'_\mathcal P\models Q_c$ (hence $Q_\mathcal P\models Q_c$) and, since $Q_c$ belongs to $\mathcal Q_j$, this contradicts the assumption that $Q_\mathcal P$ is not more specific than a CQ in $\mathcal Q_j$. 
\end{proof}

\medskip
\textbf{Proofs of the three lemmas}

\medskip
We first point out the following. 
\begin{itemize}
\item Thanks to the 
 mapping assertions $\m_{trans}$, we can always ``remove the hats" from any predicate (except the $p_{R_i}$ special predicates) in any CQ $Q_w$ belonging to a piece-rewriting of $\mathcal Q^{\qCQ,\mathcal R}$ with $\m^{\qCQ,\mathcal R}$; we just have to extend the rewriting sequence by using the rules in $\m_{trans}$. Moreover, if $Q_w$ does not contain any special atom $p_{R_i}(\vect{x_i})$, this extended rewriting is on $\mathcal S$, hence belongs to an $\mathcal S$-rewriting of $\mathcal Q^{\qCQ,\mathcal R}$ with $\m^{\qCQ,\mathcal R}$.
\item Another property of any CQ $Q_w$ belonging to a piece-rewriting of $\mathcal Q^{\qCQ,\mathcal R}$ with $\m^{\qCQ,\mathcal R}$ is that each of its terms appears in a $T$ or $\hat T$ atom. Indeed, since the CQs in $\mathcal Q^{\qCQ,\mathcal R}$ have a $\hat T$ atom for each term and all the variables of the rules in $\m^{\qCQ,\mathcal R}$ are frontier variables, no rewriting step introduces a new term without a $\hat{T}$, and the only rule that can rewrite a $\hat T$ atom replaces it with a $T$ atom.
\end{itemize}

As an immediate consequence of the previous observations, we have the following lemma.

\begin{lemma}\label{lemma-S-rewriting}
Let $Q$ be a CQ, $\mathcal R$ be a set of datalog rules and $Q_\mathcal S$ be a CQ belonging to an $\mathcal S$-rewriting of $\mathcal Q^{\qCQ,\mathcal R}$ with $\m^{\qCQ, \mathcal R}$. Then, $Q_\mathcal S$ is of form $(Q_\mathcal P)^T$, where $Q_\mathcal P$ is a CQ on $\mathcal P$. 
\end{lemma}


\medskip

\begin{lemma} \label{reduction-forward} Let $Q$ be a CQ, $\mathcal R$ be a set of datalog rules and $Q_w$ be a CQ belonging to a piece-rewriting of $Q$ with $\mathcal R$. $Q_w^T$ is isomorphic to a CQ belonging to a piece-rewriting of $\mathcal Q^{\qCQ,\mathcal R}$ with $\m^{\qCQ, \mathcal R}$.
\end{lemma}

This lemma can be proved by induction thanks to the following proposition.

\begin{proposition} \label{prop-red-forward}
Let $Q$ be a CQ, $\mathcal R$ be a set of datalog rules, $R_i \in \mathcal R$, and let $Q_w=\beta(Q,R_i,\mu)$ where $\mu$ is a piece-unifier of $Q$ with $R_i$. Let $Q_{R_i}$ and $m_{R_i}$ be respectively the CQ and $\mathcal S$-to-$\mathcal T$ rule associated with $R_i$ as defined in the reduction. There exists a disjunctive piece-unifier $\mu_\lor$ of $\{Q_{R_i}, \hat Q^T\}$ with $m_{R_i}$ such that $\hat Q_w^T$ is isomorphic to a CQ belonging to a piece-rewriting of $\{\beta_\lor(\{Q_{R_i}, \hat Q^T\},m_{R_i},\mu_\lor)\}$ with $\{T(x)\rightarrow \hat T(x)\}$.
\end{proposition}

\begin{proof}
Let $R_i=B_i[\vect{x_i},\vect{y_i}]\rightarrow H_i(\vect{x_i})$. Since $\mu$ is a piece-unifier of $Q$ with $R_i$, there is at least one atom with predicate $H_i$ in $Q$, i.e., $Q = \exists \vect{u},\vect{v}. H_i(\vect{u})\land D[\vect{u},\vect{v}]$ where $D$ is any conjunction of atoms, and $Q_w=\exists \vect{u},\vect{v},\vect{y_i}. B_i[\vect{u},\vect{y_i}]\land D[\vect{u},\vect{v}]$.
By the reduction, we obtain $\qCQ_{R_i} = \exists \vect{x_i}, \vect{y_i}. \hat {B_i}[\vect{x_i}, \vect{y_i}] \land p_{R_i}(\vect{x_i}) \land \hat T[\vect{x_i}, \vect{y_i}]$ and 
$m_{R_i} = T[\vect{x_i}] \rightarrow p_{R_i}(\vect{x_i}) \lor \hat H_i(\vect{x_i})$. We consider $\mu_\lor = \{\mu_{p_{R_i}}, \hat \mu\}$ where $\mu_{p_{R_i}}$ is the piece-unifier unifying 
 $p_{R_i}(\vect{x_i}) \in Q_{R_i}$ and $\headi{1}{m_{R_i}}$,
 and $\hat\mu$ is the piece-unifier ``isomorphic'' to $\mu$ between $\hat Q$ and $\headi{2}{m_{R_i}}$. 
More formally, given $\mu = (Q', H', P_u)$ and $\hat Q^s$ a safe copy  of $\hat Q$, $\hat\mu = ((\hat Q')^s, \hat H', P_u^s)$ where $.^s$ is the renaming function of the variables of $\hat Q$.
Clearly, the join of the partitions in $\mu_\lor$ is admissible since there is no constant. Since $\hat Q\subseteq \hat Q^T$, $\hat\mu$ is a piece-unifier of $\hat Q^T$ with the rule $T[\vect{x_i}] \rightarrow \hat H_i(\vect{x_i})$, associated with $\headi{2}{m_{R_i}}$.
Let $u$ and $u_{\mu_\lor}$ be the substitutions associated with $\mu$ and $\mu_\lor$, respectively. Then; 

$\beta_\lor(\{\qCQ_{R_i}, \hat{\qCQ}^T\}, m_{R_i}, \mu_\lor)$
$= u_{\mu_\lor}(T[\vect{x_i}] \cup \hat B_i[\vect{x_{i}}, \vect{y_i}]^s \cup T[\vect{x_{{i}}}, \vect{y_i}]^s \cup (\hat{\qCQ}^T \setminus \hat{\qCQ'})^s) 
 = u_{\mu_\lor}(\hat B_i[\vect{x_{{i}}}, \vect{y_i}]^s \cup (\hat{\qCQ} \setminus \hat{\qCQ'})^s)^T$ \\
which is isomorphic to 
 $ u(\hat B_i[\vect{x_i}, \vect{y_i}] \cup (\hat\qCQ \setminus \hat\qCQ'))^T =\hat\qCQ_w^T$.

 Since the partition associated with $\mu_{p_{R_i}}$ does not merge any frontier variables from $m_{R_i}$, no classes of $\hat \mu$ are merged in the join of the partitions of $\mu_{p_{R_i}}$ and $\hat \mu$. 
Hence, the joined partition is in bijection with $P_u^s$, and thus with $P_u$.
As a consequence, $u(\hat B_i[\vect{x_i}, \vect{y_i}] \cup (\hat\qCQ \setminus \hat\qCQ'))^T$ is isomorphic to $ u_{\mu_\lor}(\hat B_i[\vect{x_{{i}}}, \vect{y_i}]^s \cup (\hat{\qCQ} \setminus \hat{\qCQ'})^s)^T$. 
\end{proof}

\begin{proof}[Proof of Lemma \ref{reduction-forward}] 
By induction on the length $k$ of the rewriting sequence from $\{Q\}$ producing $\mathcal Q_k$ in which $Q_w$ is generated, we first prove that $(\hat Q_w)^T$ is isomorphic to a CQ belonging to a piece-rewriting of $\mathcal Q^{\qCQ,\mathcal R}$ with $\m^{\qCQ, \mathcal R}$:
\begin{itemize}
\item (k=0) $Q_w=Q$; since $Q_Q = \widehat{Q^T} \in\mathcal Q^{\qCQ,\mathcal R}$, we can produce $\hat {Q}^T = \hat {Q}_w^T$ by a rewriting sequence using the rule    
$T(x)\rightarrow \hat T(x)$. 
%

\item (k+1) 
Let $\mathcal Q_{k+1} = \mathcal Q_k \cup \{ Q_w\}$. 
Assume  $Q_w$ is generated in $\mathcal Q_{k+1}$ by a piece-unifier of $Q_k \in \mathcal Q_k$ with  $R\in \mathcal R$. 
By induction hypothesis, $(\hat {Q_k})^T$ is isomorphic to a CQ belonging to a piece-rewriting of $\mathcal Q^{\qCQ,\mathcal R}$ with $\m^{\qCQ, \mathcal R}$. Then, by Proposition \ref{prop-red-forward}, $(\hat Q_w)^T$ is isomorphic to a CQ belonging to a piece-rewriting of $\{\beta_\lor(\{Q_{R}, (\hat Q_k)^T\},m_{R},\mu_\lor)\}$ with $\{T(x)\rightarrow \hat T(x)\}$, hence a piece-rewriting of $\mathcal Q^{\qCQ,\mathcal R}$ with $\m^{\qCQ, \mathcal R}$. 

\end{itemize}
Finally, we can ``remove the hats" from any CQ belonging to a piece-rewriting of $\mathcal Q^{\qCQ,\mathcal R}$ with $\m^{\qCQ, \mathcal R}$. We just have to extend the rewriting sequence by some rewriting steps with $\m_{trans}$. Thus, since $(\hat Q_w)^T$ is isomorphic to a CQ belonging to a piece-rewriting of $\mathcal Q^{\qCQ,\mathcal R}$ with $\m^{\qCQ, \mathcal R}$, so is $Q_w^T$.
\end{proof}

\bigskip

Next, we denote by $\ruleset^{\star}$ the set of all the rules that can be obtained by composing rules from a datalog rule set $\ruleset$. Composing two datalog rules is also known as ``unfolding a rule by another''. 
Given two datalog rules $R_1= B_1 \rightarrow H_1$ and $R_2= B_2 \rightarrow H_2$, and a (most general) classical unifier $u$ of an atom $A$ in $B_2$ with the atom in $H_1$, the \emph{unfolding} of $R_2$ by $R_1$ is the rule  $R_2 \circ R_1 = u(B_1) \cup  u(B_2 \setminus \{A\}) \rightarrow u(H_2)$. Starting from $\ruleset$, one can build $\ruleset^{\star}$ by repeatedly unfolding a rule from $\ruleset^{\star}$ by a rule from $\ruleset$, until a fixpoint is reached (if any). 
 Clearly, $R_1, R_2 \models R_2 \circ R_1$. Hence, $\ruleset^{\star}$ is logically equivalent to $\ruleset$. 

\begin{proposition} \label{prop-unfolding}Let $Q$ be a CQ and $\mathcal R$ be a set of rules. Any UCQ $\qUCQ$ is a complete rewriting of $Q$ with $\mathcal R$ iff it is a complete rewriting of $Q$ with $\mathcal R^\star$.
\end{proposition}

\begin{proof}
For all fact base $F$ and CQ $Q$, one has $F, \ruleset \models Q$ iff $F, \ruleset^{\star} \models Q$. Let $\qUCQ$ be a complete rewriting of $Q$ with $\ruleset$. Then, for all $F$, $F \models \qUCQ$ iff $F, \ruleset \models Q$ iff $F, \ruleset^{\star} \models Q$, thus $\qUCQ$ is a complete rewriting of $Q$ with $\ruleset^{\star}$. Similarly,  any complete rewriting $\qUCQ$ of $Q$ with $\ruleset^{\star}$ is a complete rewriting of $Q$ with $\ruleset$. 
\end{proof}






\begin{lemma}\label{reduction-reverse}
Let $Q$ be a CQ, $\mathcal R$ be a set of datalog rules and $Q_\mathcal P$ be a CQ on $\mathcal P$ such that $(Q_\mathcal P)^T$ belongs to a piece-rewriting of $\mathcal Q^{\qCQ,\mathcal R}$ through $\m^{\qCQ, \mathcal R}$. Then, $Q_\mathcal P$ is isomorphic to a CQ belonging to a piece-rewriting of $Q$ with $\mathcal R^\star$.
\end{lemma}

To prove the lemma, we first prove some properties of the piece-rewritings of $\mathcal Q^{\qCQ,\mathcal R}$ with $\m^{\qCQ, \mathcal R}$.

\begin{proposition}\label{prop-rewriting}
Let $\mathcal Q^w$ be a piece-rewriting of $\mathcal Q^{\qCQ,\mathcal R}$ with $\m^{\qCQ, \mathcal R}$, $\mathcal Q^w$ can be partitioned into two sets: $\mathcal Q^w_M$ the subset of CQs without any $p_{R_i}$-atom, and $\mathcal Q^w_P$ the subset of CQs with exactly one $p_{R_i}$-atom. Furthermore, $\mathcal Q^w_P$ is a rewriting of $\mathcal Q_\mathcal R$ (the subset of $\mathcal Q^{\qCQ, \mathcal R}$ containing only the queries associated with the rules from $\mathcal R$), and any $\mathcal S$-rewriting of $\mathcal Q^w$ with $\m_{trans}$ is a rewriting of $\mathcal Q^w_M$ with $\m_{trans}$. 
\end{proposition}

\begin{proof}
We first show that any CQ in $\mathcal Q^w$ contains at most one $p_{R_i}$ atom:
\begin{itemize}
\item it is the case for $\mathcal Q^{\qCQ,\mathcal R}$;
\item piece-rewriting with a renaming mapping assertion in $\m_{trans}$ does not add a $p_{R_i}$ atom;
\item piece-rewriting with a disjunctive mapping assertion $m_{R_i}$ removes a $p_{R_i}$ atom (and does not add one), thus there remains at most one $p_{R_j}$ atom in the produced query.
\end{itemize}
Thus $\mathcal Q^w_M + \mathcal Q^w_P = \mathcal Q^w$.

When we use a CQ without atom $p_{R_i}$ in a piece-rewriting step, the produced query does not have such an atom either. So, we only have to consider the queries in $\mathcal Q_\mathcal R$ to generate $\mathcal Q^w_P$.

Since $p_{R_i}$ predicates do not belong to $\mathcal S$ and no rule in $\m_{trans}$ allows to rewrite a $p_{R_i}$-atom, only the queries in $\mathcal Q^w_M$ can generate queries on $\mathcal S$ using $\m_{trans}$.
\end{proof}

\begin{definition}[Reverse function]
Let $\mathcal Q^w$ be any rewriting of $\mathcal Q^{\qCQ,\mathcal R}$ with $\m^{\qCQ, \mathcal R}$. We define a ``reverse" function, noted $reverse$, from $\mathcal Q^w$ to a set of CQs plus a set of conjunctive Datalog rules, both on $\mathcal P$, as follows:
\begin{itemize}
\item for any $Q\in \mathcal Q^w_M$, $reverse(Q)=Q_r$ where $Q_r$ is the query obtained from $Q$ by removing the ``hats" on the predicates, then deleting the $T$ atoms;
\item for any $Q\in \mathcal Q^w_P$, let $Q=(\exists \vect{x},\vect{y}. p_{R_i}(\vect{x})\land C[\vect{x},\vect{y}])$. Note that $C$ is a conjunction without any $p_{R_j}$-atom. Then: $reverse(Q)=C_r[\vect{x},\vect{y}]\rightarrow H_i(\vect{x})$ where $C_r$ is the conjunction obtained from $C$ by removing the ``hats" on the predicates, then deleting the $T$ atoms, and $H_i(\vect{x})$ is obtained from the head of $R_i\in \mathcal R$ by substituting each frontier variable with the corresponding term in  $p_{R_i}(\vect{x})$.
\end{itemize}
\end{definition}

\medskip
\begin{proposition}\label{prop-P-rewriting}
Let $\mathcal Q^w$ be a piece-rewriting of $\mathcal Q^{\qCQ,\mathcal R}$ with $\m^{\qCQ, \mathcal R}$. For any $Q_w\in\mathcal Q^w_P$, $reverse(Q) \in \mathcal R^\star$.
\end{proposition}

\begin{proof}
By induction on the length $k$ of the sequence of piece-rewriting steps generating $\mathcal Q^w$ : 
\begin{itemize}
\item $(k=0)$ Recall that $\mathcal Q^w_P=\mathcal Q^{\qCQ,\mathcal R}_P = \mathcal Q_\mathcal R$. Now, observe that for each query $Q_{R_i}\in\mathcal Q_{\mathcal R}$, we have $reverse(Q_{R_i})=R_i$ which belongs to $\mathcal R$.
\item $(k+1)$ Any query $Q_w\in\mathcal Q^w_P$ is either obtained in at most $k$ piece-rewriting steps and thus $reverse(Q_w)\in \mathcal R^k$ by induction hypothesis, or there are two cases (by Proposition \ref{prop-rewriting}):
\begin{itemize}
\item 
$Q_w$ is generated by a piece-rewriting step from a CQ $Q_k$ with a $p_{R_i}$-atom and a rule in $\m_{trans}$. Then $reverse(Q_w)=reverse(Q_k)$ and, since $Q_k$ is generated in at most $k$ piece-rewriting steps, by induction hypothesis, $reverse(Q_k)\in \mathcal R^\star$.
\item $Q_w$ is generated by a piece-rewriting step from two queries $Q_1=(\exists \vect{x}_1,\vect{y}_1. p_{R_1}(\vect{x}_1)\land \hat C_1[\vect{x}_1,\vect{y}_1])$ and $Q_2=(\exists \vect{x}_2,\vect{y}_2. p_{R_2}(\vect{x}_2)\land \hat C_2[\vect{x}_2,\vect{y}_2])$ with a disjunctive rule having one of the two special predicates $p_{R_1}$ or $p_{R_2}$. Assume the rule is $m_{R_1}= T[\vect{x}] \rightarrow p_{R_1}(\vect{x}) \lor \hat H_1(\vect{x})$ associated with $R_1$. 
Let $\mu_\lor=\{\mu_1=(p_{R_1}(\vect{x}), p_{R_1}(\vect{x_1}), P_1),~\mu_2=(\hat C_2', \hat H_1', P_2)\}$ be the disjunctive piece-unifier that has produced $Q_w=u_{\mu_\lor}(T \land \hat C_1 \land p_{R_2}\land(\hat C_2 \setminus \hat C_2'))$. Then, $reverse(Q_w)=u_{\mu_\lor}(C_1 \land (C_2 \setminus C_2')) \rightarrow u_{\mu_\lor}(H_2)$.

By definition, $reverse(Q_1)=C_1 \rightarrow H_1$ and $reverse(Q_2)=C_2 \rightarrow H_2$. Let $.^s$ be the safe renaming of $Q_2$ used in $\mu_2$. 
We thus have that $\mu_2' = (C_2', H_1', (P_2)^{s^{-1}})$ is a piece-unifier between $C_2$, the body of $reverse(Q_2)$, and $H_1$, the head of $reverse(Q_1)$. It follows that $reverse(Q_2)\circ reverse(Q_1)=u_{\mu_2'}(C_1 \land (C_2 \setminus C_2')) \rightarrow u_{\mu_2'}(H_2)$.

By induction hypothesis, $reverse(Q_1)$ and $reverse(Q_2)$ belong to $R^\star$, hence $reverse(Q_2)\circ reverse(Q_1)$ belongs to $R^\star$. Since $reverse(Q_w)$ is isomorphic to  $reverse(Q_2)\circ reverse(Q_1)$, it belongs to $R^\star$. 
\end{itemize}

\end{itemize}
\end{proof}

\begin{proposition}\label{prop-M-rewriting}
Let $\mathcal Q^w$ be a piece-rewriting of $\mathcal Q^{\qCQ,\mathcal R}$ with $\m^{\qCQ, \mathcal R}$. For any $Q_w\in\mathcal Q^w_M$, $reverse(Q_w)$ is isomorphic to a CQ belonging to a piece-rewriting of $\{\qCQ\}$ with $\mathcal R^\star$.
\end{proposition}

\begin{proof} 
By induction on the length $k$ of the sequence of piece-rewriting steps generating $\mathcal Q^w$ : 
\begin{itemize}
\item $(k=0)$ $\mathcal Q^w_M=\mathcal Q^{\qCQ,\mathcal R}_M = \{Q_Q\}$ and $reverse(Q_Q)=Q$.
\item $(k+1)$ Any query $Q_w\in\mathcal Q^w_M$ is either obtained in at most $k$ piece-rewriting steps, hence, by induction hypothesis, $reverse(Q_w)$ is isomorphic to a CQ belonging to a piece-rewriting of $Q$ with $\mathcal R^\star$, or there are two cases:
\begin{itemize}
\item $Q_w$ is generated by a piece-rewriting step from a CQ $Q_k$ without $p_{R_i}$ atom and a rule in $\m_{trans}$. Then $reverse(Q_w)=reverse(Q_k)$ and, since $Q_k$ is generated in at most $k$ piece-rewriting steps, by induction hypothesis $reverse(Q_k)$ is isomorphic to a CQ belonging to a piece-rewriting of $Q$ with $\mathcal R^\star$.
\item  $Q_w$ is generated by a (disjunctive) piece-rewriting step from a CQ $Q_m=(\hat C\land \hat T)^T \in \mathcal Q^w_M$,
a CQ $Q_{R}=(\hat B_R\land p_{R_i})^T$ and the rule $m_{R_i}=p_{R_i}\lor \hat H_i$.
Let $\mu_\lor=\{\mu_i=(p_{R_i}, p_{R_i}, P_i),~\mu_2=(\hat C', \hat H_i', P_2)\}$ be the disjunctive piece-unifier that has produced $Q_w$. We thus have $Q_w=\beta_\lor(\{Q_m,Q_{R}\},m_{R_i},\mu_\lor)=u_{\mu_\lor}(\hat B_R\land(\hat C\setminus \hat C')\land \hat T)^T$, hence $reverse(Q_w)=u_{\mu_\lor}(B_R \land( C\setminus C'))$.

By Proposition \ref{prop-P-rewriting}, $reverse(Q_{R})=(B_R\rightarrow H_i) \in \mathcal R^\star$, and by induction hypothesis, $reverse(Q_m)=C$ is isomorphic to a CQ belonging to a piece-rewriting of $Q$ with $\mathcal R^\star$. 
 Let $\mu_2'$ be obtained from $\mu_2$ by replacing each predicate $\hat{p}$ with $p$ (i.e., removing the hats). 
Then, $\mu'_2$ is a piece-unifier of $reverse(Q_m)$ with $reverse(Q_{R})$ (up to a bijective variable renaming) and $\beta(reverse(Q_m),reverse(Q_{R}),\mu_2')=u_{\mu_2'}(B_R\land( C\setminus C'))$. 

With the same arguments about the join of the partitions of $\mu_\lor$ and $\mu_2'$ as at the end of the proof of Proposition \ref{prop-red-forward}, we conclude that $u_{\mu_2'}(B_R \land( C\setminus C'))$ is isomorphic to $u_{\mu_\lor}(B_R \land( C\setminus C'))$. Thus $reverse(Q_w)$ is isomorphic to a piece-rewriting of $\{\qCQ\}$ with $\mathcal R^\star$. 
\end{itemize}
\end{itemize}
\end{proof}

\begin{proof} [Proof of Lemma \ref{reduction-reverse}] 
Assume $(\qCQ_\mathcal P)^T$ belongs to a piece-rewriting of $\qUCQ^{\qCQ, \mathcal{R}}$ with $\m^{\qCQ, \mathcal{R}}$. 
Since $(\qCQ_\mathcal P)^T$ is on $\mathcal S$, it belongs to $\mathcal Q^w_M$. 
Hence, by Proposition \ref{prop-M-rewriting}, $reverse((\qCQ_\mathcal P)^T)=\qCQ_\mathcal P$ is isomorphic to a CQ belonging to a piece-rewriting of $\{\qCQ\}$ with $\mathcal R^\star$. 
\end{proof}

\end{document}